# Charge Breeding of Radioactive Ions

*F.J.C. Wenander*
CERN, Geneva, Switzerland

**Abstract**
Charge breeding is a technique to increase the charge state of ions, in many cases radioactive ions. The singly charged radioactive ions, produced in an isotope separator on-line facility, and extracted with a low kinetic energy of some tens of keV, are injected into a charge breeder, where the charge state is increased to $Q$. The transformed ions are either directed towards a dedicated experiment requiring highly charged ions, or post-accelerated to higher beam energies. In this paper the physics processes involved in the production of highly charged ions will be introduced, and the injection and extraction beam parameters of the charge breeder defined. A description of the three main charge-breeding methods is given, namely: electron stripping in gas jet or foil; external ion injection into an electron-beam ion source/trap (EBIS/T); and external ion injection into an electron cyclotron resonance ion source (ECRIS). In addition, some preparatory devices for charge breeding and practical beam delivery aspects will be covered.

## 1   Introduction and motivation

It is believed that around 6000 nuclei exist in the nuclear landscape, and presently some 3000 have been observed experimentally (of these, less than 10% are stable). The radioactive isotopes are currently the focus of nuclear physics research. In general, there are two complementary ways to generate good-quality beams of exotic nuclei, the in-flight (IF) separation method [1] and the isotope separation on-line (ISOL) technique [2]. The two methods inherently produce radioactive ions with beam energies at the opposite ends of the spectrum, IF in the GeV and ISOL in the keV region, as shown in Fig. 1. Nevertheless, the in-between region, from 0.1 to 10 MeV/$A$,[1] holds several opportunities for interesting physics [3], such as:

– Coulomb excitation;

– few-particle transfer, e.g. (d,p), ($^9$Be,2α), (p,γ); and

– fusion reactions at the Coulomb barrier.

This has pushed the technical development, and, with the introduction of post-acceleration (i.e. consecutive acceleration of ISOL-produced low-energy elements to higher energies), the energy gap could be bridged. To boost the ISOL beam energy, it is not sufficient to raise the electrostatic potential of the primary ion source (limited to a few hundred kilovolts). Therefore, conventional radiofrequency (RF) acceleration techniques such as cyclotrons or linacs have to be used. In both cases, the final beam energy per nucleon (MeV/$A$) is governed by the ion charge state: for the cyclotron with bending limit $K$, it is $K(Q/A)^2$; and for a linac with length $L_{\text{linac}}$ and average acceleration field $E_{\text{acc}}$, it is $E_{\text{acc}} L_{\text{linac}} Q/A$. The following discussion will focus on linacs, as this kind is predominantly used for post-acceleration at present-day facilities.

---

[1] Note that the term MeV/$A$, energy per nucleon, is commonly used and the unity is related to the particle velocity. The total acceleration energy is obtained by multiplying with A.

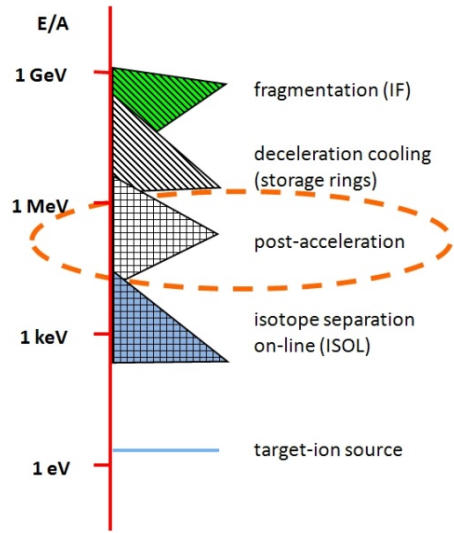

**Fig. 1:** Energy regions for nuclei produced with IF and ISOL techniques. The beam energy of IF ions can be reduced in storage rings, and the energy of ISOL beams increased in post-accelerators.

The ion charge is furthermore important for a linac as a high $A/Q$ ratio imposes a low $f_{RF}$ (a few tens of megahertz for very heavy ions with charge $1^+$) to achieve adequate

- transverse focusing (focal strength $\sim (Q/A)/f_{RF}^2$) and
- period length ($L_{period}$) of the initial RF structure part as the extraction velocity from the primary source is limited.[2]

The transverse dimension $r_{linac}$ of the cavities generally increases for lower frequencies, thus the technical design challenges and the cost increase with ion $A/Q$. In conclusion, the cost of the post-accelerator depends very much on the initial charge state, as a low $A/Q$ leads to a short linac length with small transverse dimensions, and the cost scales approximately as $L_{linac}(r_{linac})^p$, where $1 < p < 2$.

As will be seen in the following section, most ISOL systems produce $1^+$ ions. The traditional method to increase the charge of ions – the foil or gas-jet stripping technique – was challenged some 15 years ago by novel schemes for charge breeding. The novelty was the transformation from $1^+$ to $Q$ charged ions inside an electron-beam ion source/trap (EBIS/T) or electron cyclotron resonance ion source (ECRIS) by electron–ion collisions. These charge breeders are located in the low-energy part of the machine before the accelerating structures. Because of the capability of these charge-breeding devices to produce highly charged ions, $A/Q$ ratios between 3 and 9 are easily obtained. The layout of a combined ISOL and post-accelerator system is shown in Fig. 2, including an $A/Q$ analyser after the breeder, which selects the correct $A/Q$ value and separates the radioactive ions from stable ion beams formed from the residual gas.

## 2  ISOL beam parameters and breeder criteria

The charge breeder acts a link between the ISOL production stage and the post-accelerator. It has to accept the different beams effectively and deliver charge-bred beams with the right characteristics to the post-accelerator. The properties of the radioactive ion beams from an ISOL facility may vary greatly depending on the element, type of ion source and if preparatory devices are being used or not.

---

[2] For example: $A = 220$, $Q = 1$, $U_{extr} = 100$ kV $\Rightarrow v_{extr} = 3 \times 10^5$ m s$^{-1}$; $L_{period} = 2$ cm $\Rightarrow f_{RF} \sim v_{extr}/L_{period} = 15$ MHz. Unless otherwise indicated SI units are used throughout the paper.

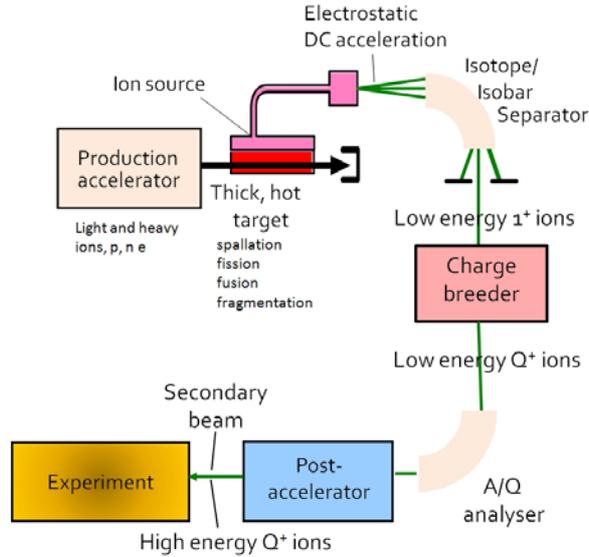

**Fig. 2:** Schematic layout of an ISOL production system with subsequent charge breeder and post-accelerator, followed by an experimental set-up.

## 2.1   What goes in …

The ion mass stretches from light $A = 6$ (even $^4$He is used for test purposes) to very heavy elements with $A \sim 238$. As the neutron drip-line is further away from stability than the proton drip-line, many of the ions have a large neutron excess. This means that, to reach the required $A/Q$ ratio, charge breeding to a relatively high charge state is imposed. The ion current spans an extensive region from just a few ions per second to $>10^{11}$ ions per second, where the upper value may increase with next-generation ISOL facilities. The ion charge is predominantly $1^+$ but sometimes $2^+$ or even higher charges can be produced by the initial ion source on the target and transferred to the charge breeder. The beam energy is a few tens of keV (total energy) with an energy spread of a few eV. If the driver beam is continuous, the radioactive ions are released continuously from the primary target; although, with a pulsed driver, a semi-continuous, but intensity-modulated, time structure is introduced. The release time, ranging from some tens of milliseconds to minutes or longer, is mainly governed by the diffusion of the radioactive atoms in the target, the transport process from the target to the ion source, and the half-life of the ion itself. An example of a time release curve for $^8$Li is given in Fig. 3.

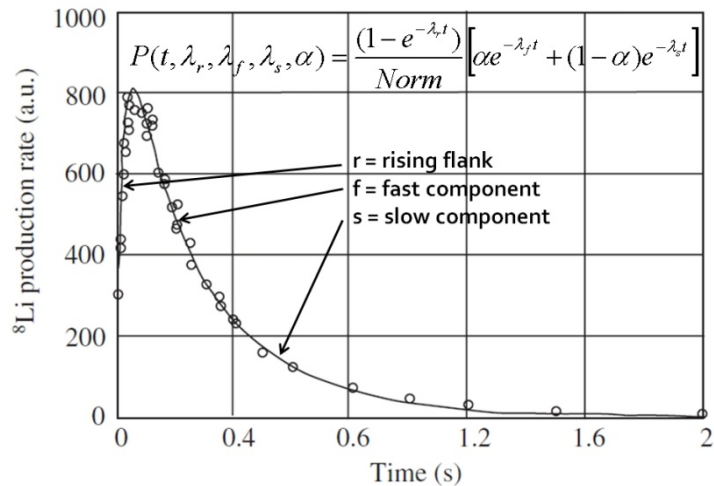

$$P(t,\lambda_r,\lambda_f,\lambda_s,\alpha) = \frac{(1-e^{-\lambda_r t})}{Norm}\left[\alpha e^{-\lambda_f t} + (1-\alpha)e^{-\lambda_s t}\right]$$

r = rising flank
f = fast component
s = slow component

**Fig. 3:** Release curve of $^8$Li ions from a tantalum foil target. A proton pulse impinges on the target at t = 0 s. An empirical fitting formula is included. Based on [4].

When one moves away from the valley of stability, the half-life of the nuclide to be investigated is typically reduced. Close to the drip-lines it can be as short as a few milliseconds or even shorter. Nevertheless, the production and ionization processes in the ISOL system set a lower limit of around 10 ms. Finally, the extracted beam might not be isobarically clean but might have a superposition of isobaric contaminants. They can be abundant, even dominant, particularly far from stability. Several techniques to suppress them within the ISOL system exist, such as using resonant laser ionization [5], chemical suppression and molecular sideband beams (see section 7.2), to mention but a few. The transverse emittance of the $1^+$ radioactive beam is strongly linked to the type of ion source (e.g. plasma or surface ionizer) and its design. The beam properties of the ISOL system are summarized in Table 1.

**Table 1:** Summary of ISOL beam parameters.

| Entity | Value | Comments |
| --- | --- | --- |
| Ion mass | 4 to ~238 | He to U |
| Intensity | Few to >$10^{11}$ ions/s | Large dynamic range |
| Charge | $1^+$ | Sometimes $2^+$, $3^+$, … |
| Energy | Several tens of keV | Total beam energy |
| Energy spread | Few eV | |
| Temporal structure | CW or quasi-CW | Driver beam – CW or pulsed |
| Transverse emittance | 10–50 mm mrad | 90% at 50 keV |
| Half-life | >10 ms | Limited by ISOL system |
| Selection | Not necessarily isobarically clean | Make use of resonant laser ionization, for example |

## 2.2   … and what comes out

The charge breeder has to produce multiply charged ions. Extremely high charge states are not required in most cases, as long as the upper $A/Q$ limit of the accelerator is fulfilled (typically between 4.5 and 9), although exceptions exist, as will be pointed out below. The breeding efficiency $\eta_{\text{breed}}$, defined as the number of extracted divided by the number of injected particles, or expressed in terms of electrical currents as

$$\eta_{\text{breed}} = \frac{I_{\text{extr}}(Q)}{Q I_{\text{inj}}(1^+)}, \qquad (1)$$

where $I_{\text{inj}}$ and $I_{\text{extr}}$ are the injected and extracted electrical currents, is of utmost importance. At least 10% is expected, as many isotopes are rare and difficult to produce. Moreover, a high efficiency leads to reduced machine contamination. To further reduce the losses, the breeding time should be short (ideally down to 10 ms) for short-lived ions. Apart from these criteria, a sufficient throughput capacity should be provided for high-intensity beams (up to microamps), some serving as secondary driver beams for the production of even more neutron-rich nuclei or superheavy elements. Another important factor is the purity of the extracted beam, as stable beam contaminants from the breeder may disturb the experiment, as will be described later. Ideally, the breeder system should suppress isobaric and molecular contaminants coming from the incoming ISOL system. To trap the injected ions efficiently, the transverse acceptance should be large, covering the emittance of the ISOL beam. The extracted beam emittance, on the other hand, should be as small as possible to achieve a sufficient resolution in the subsequent $A/Q$ separator and to fit into the acceptance space of the accelerator. It is possible that the output emittance of the beam from the charge breeder is smaller than the incoming beam emittance, owing to the cooling effect of the ions in the charge breeder plasma. Depending on the post-accelerator and experiment, a continuous wave (CW) or pulsed extracted beam is requested. For low intensities, a pulsed beam at the experiment might be preferable to increase the signal-to-noise ratio, while high-intensity beams should be delivered as CW in order to avoid detector saturation.

## 3    Atomic physics processes for multiply charged ions

The removal of an electron from an atom or molecule requires an electric field in excess of $10^{10}$ V m$^{-1}$ for field ionization, only achievable within atomic distances typically reached in collisions with charged particles (or intense lasers). The conservation of energy and momentum favours electrons as the most efficient ionizing particles compared to protons and photons, for example. The electron-impact ionization cross-sections $\sigma$ are of the order of $10^{-16}$ cm$^2$ for neutral atoms, approximately matching the geometrical size of the atoms. There are two different ways of producing multiply charged ions:

–  in a single collision, where many electrons are removed from the ion (double ionization, triple ionization, etc.); and

–  by multistep ionization, where only one electron is removed per collision and high charge states are produced in successive different collisions.

In single-step ionization, the incident electron must have an energy of at least the sum of all the binding energies of the removed electrons, whereas in multistep ionization they only have to exceed the energy of each electron removed. In practice, multistep ionization, expressed as

$$\mathrm{e}^- + \mathrm{A}^{i+} \rightarrow \mathrm{A}^{(i+1)+} + 2\mathrm{e}^- \qquad (2)$$

is the only feasible route to high-charge-state ions, but the process takes time. The time depends on the plasma density and the ionization cross-section, and must be shorter than the ion lifetime in the plasma, limited by the confining field(s) and recombination processes. The mean time $\bar{\tau}_Q$ required to reach a charge state $Q$ can be calculated by adding up the individual ionization times, assuming no multiple ionization:

$$\bar{\tau}_Q = \sum_{i=1}^{Q-1} \bar{\tau}_{i \rightarrow i+1} = \frac{e}{j_\mathrm{e}} \sum_{i=1}^{Q-1} \frac{1}{\sigma_{i \rightarrow i+1}}. \qquad (3)$$

According to the expression, valid for a monoenergetic electron beam, the breeding time can be adjusted by varying the electron current density $j_\mathrm{e}$ (A cm$^{-2}$).

The ionization cross-section can be estimated using Lotz's semi-empirical formula [6]. The simplified version given in Eq. (4) is valid for electron-impact beam energies $E_\mathrm{kin}$ much larger than the electron binding energies $E_{i,nl}$,

$$\sigma_{Q \rightarrow Q+1} = 4.5 \times 10^{-14} \sum_{nl} \frac{\ln(E_\mathrm{kin}/E_{i,nl})}{E_\mathrm{kin} E_{i,nl}} \quad (\mathrm{cm}^2), \qquad (4)$$

where $E_{i,nl}$ denotes the binding energy for electron $i$ in subshell $nl$, the energies are given in eV and the sum is over all removable electrons in all orbitals $nl$. This typically leads to a maximum of the ionization cross-section when the electron energy is approximately three times the ionization potential. The Lotz formula gives approximate values with uncertainty of +40%/−30% for most species [7]. The formula is based on rather low-energy experimental data and should be applied with care for highly charged ions of high $Z$. For these cases, one should consider use of atomic numerical codes such as Flexible Atomic Code [8], paying attention to relativistic treatment for the highest ionization energies.

Ionization is a statistical process leading to a charge-state development as seen in the upper part of Fig. 4. The width of the distribution given in the lower part of the figure depends mainly on the $Z$ of the element, and usually 15–35% of the ions are found in the most abundant charge state. If the electron distribution is monoenergetic, its energy can be chosen just below an ionization level, thereby pushing a very high fraction of the ions into more or less a single charge state.

Electron-impact ionization is not the only process on-going inside a plasma. Several competing processes reduce the ion charge, for instance radiative and dielectronic recombination with electrons, and charge exchange with other atom/ion species. In addition, the ions may escape from the confinement. A differential equation describing the evolution of one charge state is given in Eq. (5). The model is suitable to be used for ions in an EBIS/T, where the electron-beam energy is known, partial neutral gas pressure can be estimated and the trapping potential easily calculated:

$$\frac{dN_i}{dt} = n_e v_e \left[ \sigma^{EI}_{i-1 \to i} N_{i-1} - \left( \sigma^{EI}_{i \to i+1} + \sigma^{RR}_{i \to i-1} + \sigma^{DR}_{i \to i-1} \right) N_i + \left( \sigma^{RR}_{i+1 \to i} + \sigma^{DR}_{i+1 \to i} \right) N_{i+1} \right] \\ - n_0 v_{ion} \left[ \sigma^{CX}_{i \to i-1} N_i - \sigma^{CX}_{i+1 \to i} N_{i+1} \right] - N_i R^{ESC}_i. \quad (5)$$

Here $N_i$ denotes the number of ions with charge $i$, $n_e$ and $v_e$ the electron density and velocity, $n_0$ the neutral particle density of other atoms and $v_{ion}$ the average ion velocity, defined by the ion injection energy, ionization and electron scattering heating [10]. The index EI denotes electron-impact ionization, RR radiative recombination, DR dielectronic recombination, CX charge exchange and ESC escape rate. CX between different ion charge states is suppressed due to Coulomb repulsion of ions, and therefore only charge exchange with neutrals is considered.

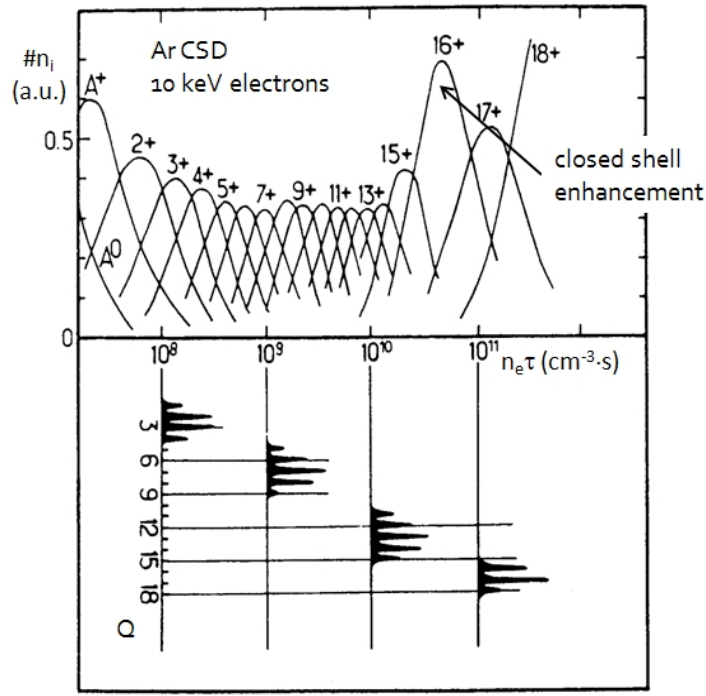

**Fig. 4:** Calculated relative atom and ion density of argon under bombardment of monoenergetic 10 keV electrons as a function of electron density times breeding time. The calculation assumes only single-step ionization. The He-like properties of $Ar^{16+}$ reduce the cross-section for ionization from $16^+$ to $17^+$ compared to $15^+$ to $16^+$ and therefore enhances the abundance of $16^+$. Based on [9].

The charge exchange process can be calculated using the Müller–Salzborn empirical formula [11] or the Selberg approximation [12]. The significance of charge exchange is defined by the ratio of EI and CX probabilities at given electron and neutral densities, electron and ion velocities and corresponding cross-sections, and is of little importance for low ion charge states.

For instance, for $Pb^{20+}$ immersed in an electron-beam current density of 100 A cm$^{-2}$, the rate of electron-impact ionization equals the charge exchange rate with neutrals at a relatively high pressure of a few $10^{-7}$ mbar. For $Pb^{70+}$, on the other hand, the rates are similar at a significantly lower pressure of $10^{-11}$ mbar. Likewise, during the breeding to high charges, the ions gain significant momentum from elastic scattering by the highly energetic electrons, making them more likely to be lost from the confinement potential, and as a result of the higher ion velocity the CX rate is increased. The increase in ion charge state, however, augments the confinement force of the electric fields, and ion cooling actions can be present, both processes reducing the ion loss rate. Finally, to minimize the radiative recombination, the energy of the electrons should be considerably higher than the ionization potential, although that lowers the electron-impact ionization cross-section. For highly charged ions, the final population of the highest charge states is defined by the equilibrium between RR and EI at the given electron density and velocity. This is illustrated for the uranium case in Fig. 5, where the two graphs plotted indicate the residual gas pressure at which the cross sections for EI and CX, and RR and CX processes, are equal. Compared to the lead example above, a higher residual gas pressure can be tolerated as the electron current density is significantly higher.

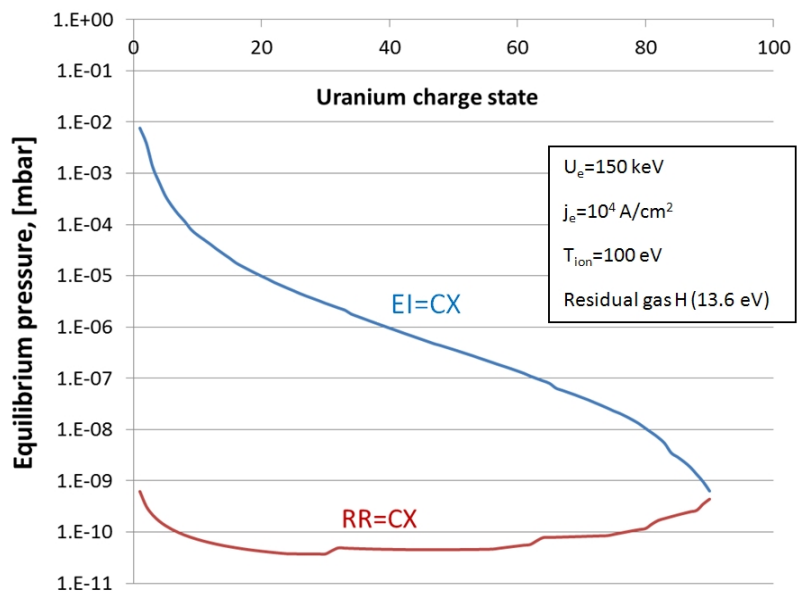

**Fig. 5:** The two graphs plot the residual gas pressure at which the cross sections for EI and CX, and RR and CX equal for charge breeding of uranium in a very high performance EBIS/T.

## 4 The gas-jet/foil stripper alternative

Although strictly not considered as a charge breeder, charge stripping in a dense, cold electron cloud is the classical way of increasing the charge state. This method is used at numerous heavy-ion injectors, for example, in the Pb injector chain at CERN. The idea of shooting ions with low charge through either a thin foil or a layer of gas in order to strip away further electrons has been used more or less since the invention of ion accelerators, and will therefore be introduced briefly as a reference for the other breeder types.

The ions need to be pre-accelerated to around 10 keV/*A* for efficient stripping in a gas jet, and up to several 100 keV/*A* for stripping in a solid foil. The stripping process may be repeated at different ion velocities in order to reach even higher charge states, as will be explained below. Usually He is used as gas-jet stripper medium, and Be, C, Al, $Al_2O_3$ or Mylar are used as foil material. Carbon foils have the advantage of being stable in vacuum at high temperatures, combined with good electrical and thermal conductivity. Furthermore carbon has the advantage of being the material with the lowest *Z* that can be fabricated into very thin foils to minimize multiple scattering and energy straggling [13].

The foil contains in its crystalline structure cold electrons with a density $n_e$ of around $10^{24}$ cm$^{-3}$. The relative interaction velocity equals approximately the transit velocity $v_{proj}$ of the accelerated ions ($10^8$ to a few $10^9$ cm s$^{-1}$). The interaction time $t_{transit}$, of the order of $10^{-14}$ s, is given by the transit time, thus $n_e t_{transit} \sim 10^{10}$ s cm$^{-3}$ and $n_e v_{proj} t_{transit} > 10^{19}$ cm$^{-2}$. During the interaction, two types of collisions are in competition: step-by-step ionization, and recombination through electron capture. At high speed, the former dominates. According to Bohr's criterion, an ion penetrating through matter retains only those electrons whose orbital velocity is greater than the ion velocity $v_{proj}$. Baron's empirical formula [14] can be used to calculate the mean equilibrium charge state $\bar{Q}$ for carbon foil stripping:

$$\bar{Q} = Z_{proj} C_1 \left( 1 - C_2 \exp\left( \frac{-83.28\beta}{Z_{proj}^{0.447}} \right) \right), \quad (6)$$

where relativistic $\beta = v_{proj}/c$, $C_1 = 1$ for $Z_{proj} < 54$, $C_1 = 1 - \exp(-12.9 + 0.21 Z_{proj} - 0.001 Z_{proj}^2)$ for $Z_{proj} \geq 54$, and $C_2 = 1$ for ion energies $W_{proj} \geq 1.3$ MeV/$A$, $C_2 = 0.9 + 0.08 W_{proj}$ for $W_{proj} < 1.3$ MeV/$A$. Note that the average charge state is only dependent on $v_{proj}$ and $Z_{proj}$. In Fig. 6 the attainable charge states as a function of the ion number for different particle velocities are plotted.

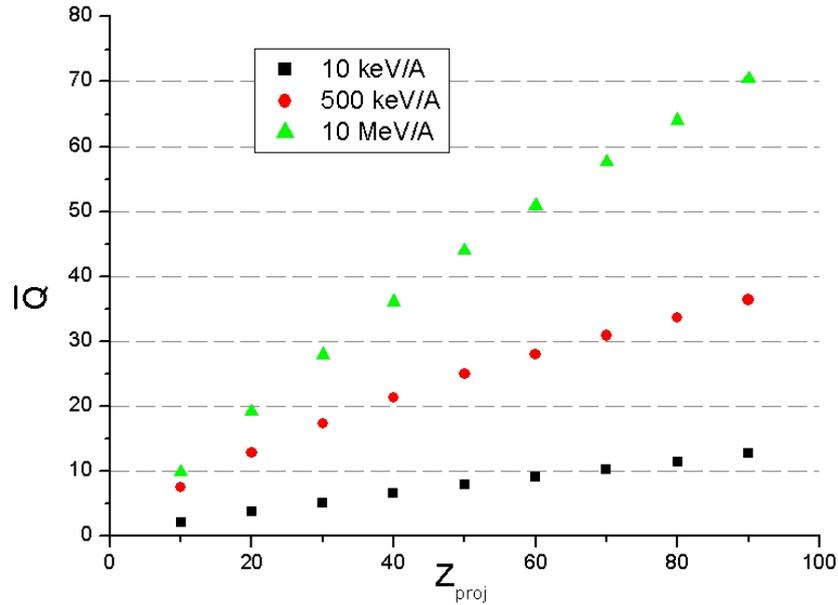

**Fig. 6:** Mean equilibrium charge state as a function of ion number for three different projectile energies after stripping in a carbon foil.

The charge-state distribution can be assumed to be Gaussian if no significant atomic shell effects are present and $\bar{Q}$ is not too close to $Z_{proj}$. The width $\sigma_w$ of the distribution is given by

$$\sigma_w = 0.5 \sqrt{\bar{Q} \left[ 1 - \left( \frac{\bar{Q}}{Z_{proj}} \right)^{1.67} \right]} \quad \text{for } Z_{proj} < 54, \quad (7a)$$

$$\sigma_w = \sqrt{\bar{Q} \left[ 0.075 + 0.19 \left( \frac{\bar{Q}}{Z_{proj}} \right) - 0.27 \left( \frac{\bar{Q}}{Z_{proj}} \right)^2 \right]} \quad \text{for } Z_{proj} \geq 54. \quad (7b)$$

Light elements will have a narrow distribution, leading to a relatively high fraction in a single charge state, while heavy elements will have the ions distributed over more charges, as demonstrated in Fig. 7. Revised formulas for the stripping processes in foils and gases are given in Ref. [15].

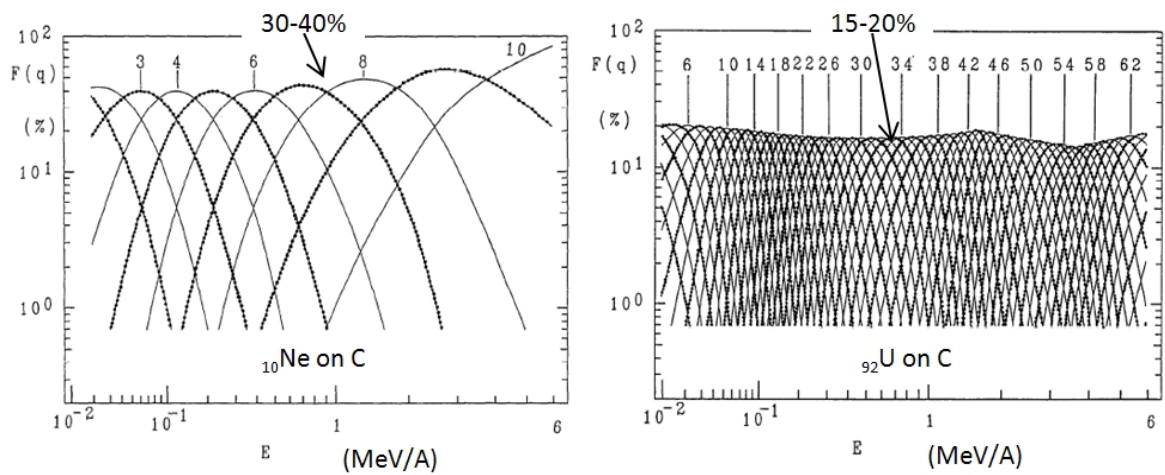

**Fig. 7:** Equilibrium charge-state distributions for Ne (left) and U (right) after stripping in a carbon foil. The plots show the fraction in different charge states as a function of the projectile energy. Based on [16].

The discussion above assumes that an equilibrium charge-state distribution has been reached, which means that the distribution at the stripper exit does not depend on the initial ion charge-state distribution, and nor does it change if the thickness of the target is further increased. It is important to know the equilibrium thickness, as the desired charge state is not attained if the foil is too thin, and excessive thickness will introduce further energy straggling and loss, and angular straggling.[3] From the approximation in Fig. 8, we learn that the carbon foil equilibrium thickness for a 500 keV/$A$ beam is < 5 µg cm$^{-2}$ (corresponding to < 25 nm). Manufacturing and handling of thinner foils is practically very difficult, and therefore gas targets are used for lower particle velocities.

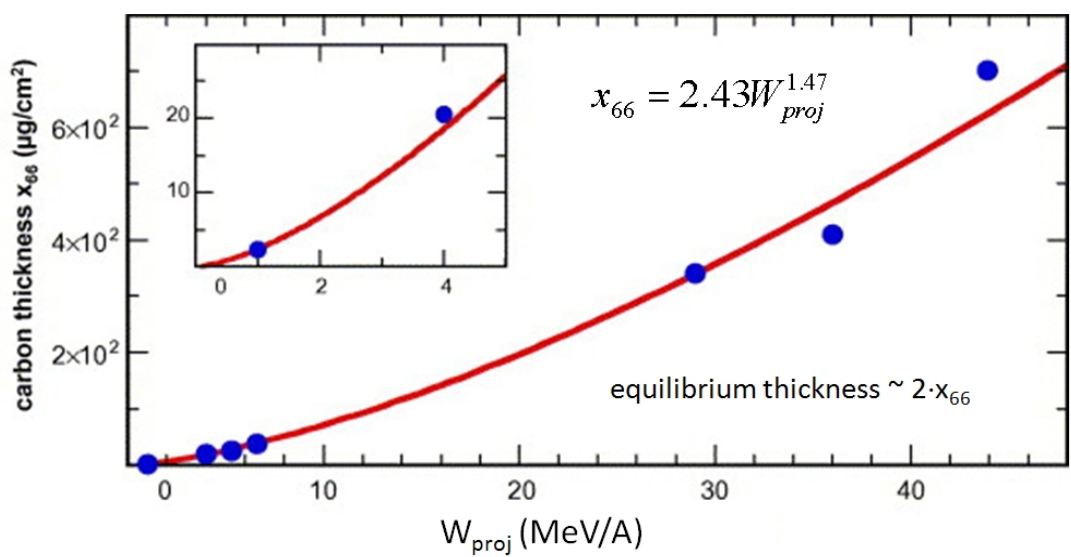

**Fig. 8:** Approximate equilibrium thickness for carbon stripper foil as a function of projectile energy. Based on [17].

The integrated thickness for a gas-jet stripper is a fraction of a microgram per square centimetre, suitable for ion velocities between 5 and 25 keV/$A$. Even though a relatively high fraction can be attained in a single charge state (35–55%), the charge is modest, ranging from $2^+$ to $4^+$ when going from Xe to U. It is interesting to note that in a solid stripper the collision frequency is larger than in a dilute gas-jet stripper, leading to less pronounced recombination effects due to Auger and radiative decays, and therefore a higher average charge for the same thickness.

---

[3] Calculated using, for example, the SRIM/TRIM package [18].

As the attainable charge in a stripping process is velocity-dependent, one would ideally like to introduce a stripper stage as soon as the increased particle velocity enables a higher charge state in order to make best use of the acceleration voltage. This would, however, lead to poor transmission due to the selection of a single charge state after each stripping stage. Instead, a typical layout makes use of one gas stripper followed by one or two stripper foils. See Table 2 for typical numbers on post-acceleration of $^{132}$Sn using two stripper stages.

**Table 2:** Charge stripping of $^{132}$Sn.

| Entity | Transmission | Comments |
| --- | --- | --- |
| Bunching efficiency[a] | 65% | |
| Gas stripping to $2^+$ | 55% | At 8 keV/$A$ |
| Stripping foil to $23^+$ | 20% | At 500 keV/$A$ |
| Total transmission | 7% | In a single charge state |

[a] A bunch-rotating RF cavity is mandatory in order to generate a time focus at the stripper to minimize the longitudinal emittance growth due to energy straggling.

To increase the efficiency for heavy-ion beams, where multiple stripping stages are required, multi-charge-state acceleration is an option. Instead of selecting a single charge state after the stripping, a broader band of charges is accelerated, leading to a significant increase (a factor 2–3) in particle intensity for high-$Z$ ions [19]. A charge variation acceptance $\Delta Q/Q \sim 10\%$ can be achieved in a linac. The disadvantages are complicated beam transport and an increased transverse and longitudinal emittance of at least a factor of 3 compared with single-charge-state acceleration. Precise longitudinal beam manipulations are also required, with an additional cavity for phase synchronization after each stripping station, and another one for beam re-collection before each additional stripping station [20].

To summarize, the foil (gas-jet) stripping method has several attractive features:

– It is simple, as it only uses passive elements and the foil arrangement itself is inexpensive.
– The ionization process equals the travel time through the foil, which is sub-picosecond.
– It is capable of handling very high beam intensities.
– It introduces no extra beam contamination to the radioactive beam.
– It is a very efficient method for the production of bare light ions.

The method has, however, a few drawbacks, summarized below:

– It needs pre-acceleration in order to attain sufficient velocity for stripping in either gas-jet or thin solid foil. The cost of the pre-accelerator adds to the overall project cost.
– The quasi-CW macrostructure[4] of the beam, determined by the release properties of the ISOL system, calls for a CW post-accelerator in order not to introduce losses.
– The 6D phase-space beam emittance increases due to multiple angular scattering and energy-loss straggling of the ions in the stripper material.
– The lifetime of the carbon foils is limited. Nevertheless, for most radioactive beams, the foil lifetime should be of little concern, as the intensity is low. It is not sufficient to reach sublimation temperatures (>150 W cm$^{-2}$), and a long operation time before radiation damage occurs is expected.

---

[4] A microstructure (several MHz) is imposed on the beam by the pre-buncher and the RF accelerating cavities.

# 5 ECRIS-based charge breeders

The concept of using an ECRIS as charge breeder is straightforward, as illustrated in Fig. 9. Singly charged ions are injected slowly into an ECRIS, and subsequently captured by the plasma, where hot electrons perform electron-impact step-by-step ionization of the ions. The highly charged ions migrate out of the plasma and are extracted by an extraction electrode before being further post-accelerated.

The ECRIS has to be of minimum $B$ (min-B) field type, which means that the magnetic field increases in all directions from the centre of the magnet configuration, and the plasma is confined longitudinally by Helmholtz coils and radially by a strong permanent magnet multipole.

The electrons are confined by the magnetic field and the ions in addition through the charge-neutrality condition of the plasma. In a min-B field device, the ions can be confined for a longer time (some 100 ms) and high-energy electrons (exceeding 100 keV) are present.

In Fig. 10 the required electron energy and electron density multiplied by confinement time are indicated for different ion charges. As the confinement time is around 0.1 s, an electron density $n_e \sim 10^{12}$ cm$^{-3}$ is necessary for charge breeding. The plasma cut-off frequency, that is, the lowest frequency that can penetrate a plasma with a certain electron density, is around 10 GHz, given by Eq. (8) – up to now ECRIS breeders have usually operated at a frequency of 14–18 GHz:

$$f_p = \frac{1}{2\pi}\sqrt{\frac{e^2 n_e}{\varepsilon_0 m_e}}. \qquad (8)$$

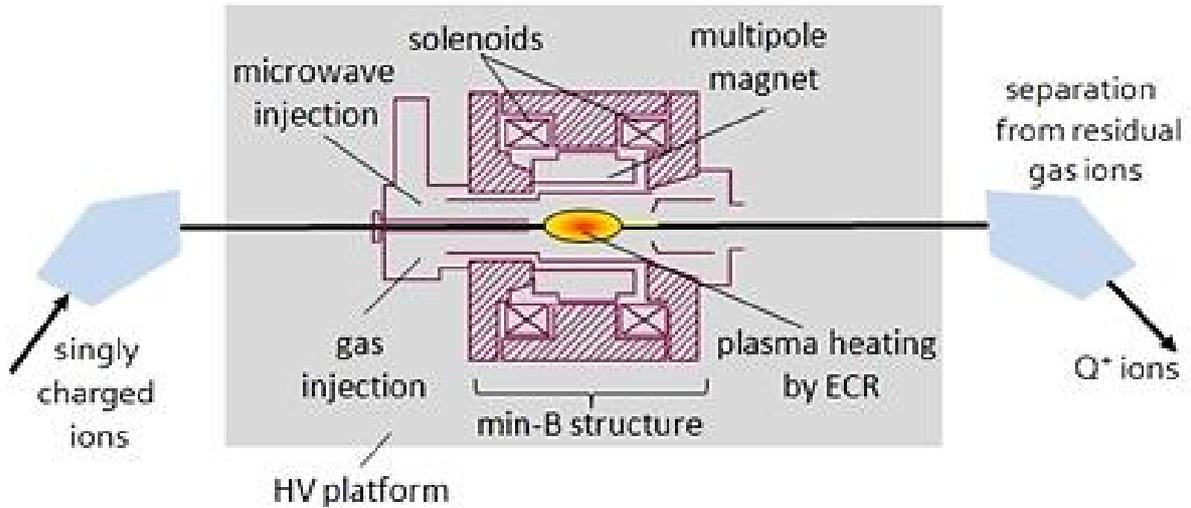

**Fig. 9:** Schematic concept making use of an ECRIS for charge breeding

Here $m_e$ and $e$ represent the mass and charge of the electron, respectively. As the electron density and confinement time are given, the breeding capacity and ion throughput can be estimated. For this we assume a plasma volume of 100 cm$^3$ and charge breeding to 10$^+$. Approximately 20% of the ions are found in charge state 10$^+$, with the rest in the neighbouring charge states.

Finally we assume that the radioactive ions occupy 10% of the total space charge, the rest being support gas ions. A total of $2 \times 10^{12}$ radioactive ions per second would then be extracted, corresponding to 400 particle nanoamps. The estimation has been confirmed with tests where beam intensities of a few microamps have been injected and successfully charge-bred [21]. Such a large capacity is an important strength of the ECRIS charge breeder concept.

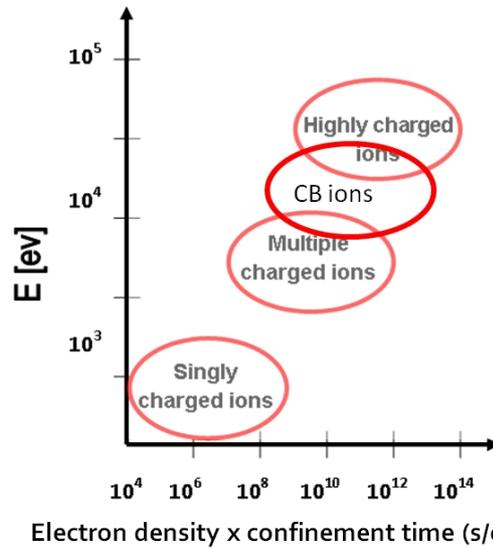

**Fig. 10:** Required electron energy and density multiplied by confinement time for production of different ion charge states.

The ions entering an ECRIS charge breeder are not trapped in a strong electrostatic potential from an electron beam, as is the case inside an EBIS/T. Instead, the singly charged ions are first slowed down electrostatically by positioning the ECRIS on a positive high-voltage potential, then they are thermalized inside the plasma region, and finally ionized to a higher charge state and thereby trapped by a weak electrostatic potential dip in the centre of the plasma. The general stopping procedure, the so-called plasma capture, is depicted in Fig. 11.

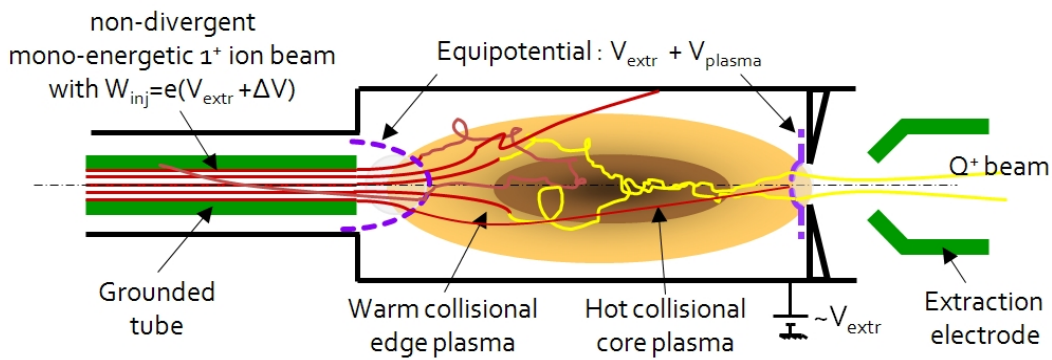

**Fig. 11:** The stopping process, or plasma capure, for injection of $1^+$ ions into an ECRIS plasma. After trapping and stepwise ionization, the multiply charged ions subsequently migrate towards the extraction side and are accelerated by the extraction electrode.

An ion entering a plasma is deflected by the Coulomb force from the ions and electrons. In a strongly ionized, quiescent plasma, as found in a min-B structure ECRIS, the cumulative deflection due to small-angle, long-range scattering with many particles is larger than single two-body large-angle scattering. Each collision with a plasma electron or ion only gives a small-angle deviation to the injected particle, but the cumulative effect of many small-angle collisions eventually yields large-angle scattering of 90°. During the collision process, energy is transferred from the injected particle to the plasma charges.

In the complete collision damping theory proposed by Spitzer [22], the incident particles with velocity $v_{inj}$ transfer momentum via collisions with the thermal electrons and ions of the magnetoplasma with velocities $v_{therm-}$ and $v_{therm+}$, respecitecely. For $v_{inj}/v_{therm+} < 10$ the ion–ion collision is the guiding mechanism, while for higher ratios the damping efficiency drops rapidly until the ratio $v_{inj}/v_{therm-}$ reaches values of the order of unity. The collisions between injected ions and the thermal electrons then become frequent. However, the mean free path length for this type of collision is too long to play a role in an ECRIS plasma. Thus, the governing slowing-down process is ion–ion collisions, and the slowing-down coefficient can be calculated with [23]

$$\frac{\langle \Delta v_{inj} \rangle}{\Delta t} = \frac{n_{therm+}}{2\pi\varepsilon_0} \left[ \frac{Q_{inj}Q_{therm+}e^2}{m_{inj}\overline{v}_{therm+}} \right]^2 \left(1 + \frac{m_{inj}}{m_{therm+}}\right) R\left(\frac{v_{inj}}{\overline{v}_{therm+}}\right) \ln \Lambda . \quad (9)$$

Here $n_{therm+}$ is the density of plasma ions, $Q_{inj}$ and $Q_{therm+}$ the charge states of the injected particle and the mean charge state of the plasma particles, respectively, $m_{inj}$ and $m_{therm+}$ their masses, $\ln \Lambda$ the Coulomb logarithm and $R(v_{inj}/\overline{v}_{therm+})$ a function of the injected particle and plasma ion velocities. We see that high plasma density and high ion charge states lead to a rapid slowing down. This has also been proven, as doubly charged Pb ions are easier to trap than singly charged. The function $R(v_{inj}/\overline{v}_{therm+})$ peaks when the velocities are similar, which has some consequences for the ion injection [24]. For example, if $^{85}$Rb$^+$ ions enter a plasma mainly constituted by $^{16}$O ions from the support gas at a temperature of approximately 2 eV, their energy should ideally be 2 eV times 85/16, which equals ~10 eV. Ions with much higher injection energy, for instance 40 eV, would not be trapped. Following this argument, the injection energy of light ions should be low, around 2 eV for $^{11}$Li$^+$, which is challenging from a beam optics and magnetic stray-field point of view.

It should be noted that a few assumptions are essential for the capturing model to be valid, namely, the following:

– The intensity of the injected particles is low, meaning that they do not interact with each other. In reality, this means that the injected ion intensity should only be a fraction of the support gas ions inside the plasma.
– The injected particles interact with the plasma particles only via long-range cumulative plasma collisions.
– The plasma particles have a Maxwellian velocity distribution.
– The distance between 90° deviations is smaller than the plasma size.

Is the critical condition for 90° deflection inside an ECRIS fulfilled? The mean free path for a 90° deviation, $\lambda_{90°}$, can be estimated with the following expression [25]:

$$\lambda_{90°} \approx \frac{W_{inj}}{4\pi n_e Q_{inj} Q_{therm+} e^2 \ln \Lambda} , \quad (10)$$

where $W_{inj}$ is the energy of the injected particle inside the plasma. If we assume $W_{inj} = 10$ eV (as shown above to be reasonable), $Q_{inj} = 1$, $Q_{therm+} = 10$, $n_e = 10^{12}$ cm$^{-3}$ and approximate $\ln \Lambda$ with 10, then we find that $\lambda_{90°} \sim 5$ cm, thus comparable with the dimensions of the ECRIS plasma. The thermalization process of the injected ions takes a few ion–ion plasma collisions, and within some $10^{-4}$ s they have obtained a Maxwellian temperature distribution with an energy of some eV and are then confined like the support gas ions. Subsequently, the energetic electrons can strip the 1$^+$ ions through step-by-step ionizing collisions and transform them into higher-charge-state $Q$ ions.

Based on the theory for the ion capture inside the plasma, one can conclude that the high-voltage adjustment of the ISOL ion source relative to the charge breeder is crucial. The layout of the electrical potentials is shown in Fig. 11 (left) and due to the often unknown plasma potential (i.e. the potential difference between the plasma and the plasma chamber; typically some tens of eV) inside the breeder, the voltage setting needs to be tuned. The normalized extracted charge-bred current is plotted as a function of the charge breeder potential for different primary ion sources in Fig. 12 (right). If the charge breeder potential is too low, meaning that the ions enter the plasma with a high velocity, only noble gases are charge-bred. These ions can impinge at the plasma chamber wall without sticking and thereafter re-enter into the plasma region. Therefore the energy range of non-sticking elements is large (some 100 eV). Condensable, or metallic, ions, on the other hand, need to have the exact entrance energy (within some eV): if the energy is too low, they will not enter the plasma region; and if too high, they will hit the chamber wall and be stuck. The mean stay time (sojourn time) for atoms on the wall, given by Frenkel's law, is for Ar only some 10 ps, while for Ni it is around 100 years at room temperature.

The ability to adjust the $A/Q$ value of the radioactive ion is essential in order to avoid superimposed stable beam contaminants. In an ECRIS the charge state is mainly dependent on the electron density $n_e$, the ion confinement time $\tau_{ion}$ and the electron energy distribution. These properties cannot be adjusted directly, so instead the accessible parameters are tuned. In practice, the injected RF power, and thus the power density, is varied as $Q \propto P_{RF}^{1/3}$. Secondly, the support gas pressure, which influences the number of cold electrons, the ion–ion cooling effect and charge exchange probability, can be changed. Finally, the magnetic field at the extraction side, $B_{ext}$, is varied as $Q \propto \ln(B_{ext})$. The latter effect can be explained by the magnetic-mirror effect, consisting in reflecting the charged particles back towards the weak-field region (illustrated in Fig. 13) with a force

$$F = -\frac{1}{2}\frac{mv_\perp^2}{B}\frac{dB_z}{dz} = -\mu\frac{dB_z}{dz},$$

where $v_\perp$ is the ion velocity perpendicular to the axial $B$ field and $\mu$ the magnetic moment of the ion. The axial confinement is due to the conservation of total kinetic energy and the magnetic moment. A higher magnetic field reduces the loss rate of ions through the loss cone of the magnetic bottle.

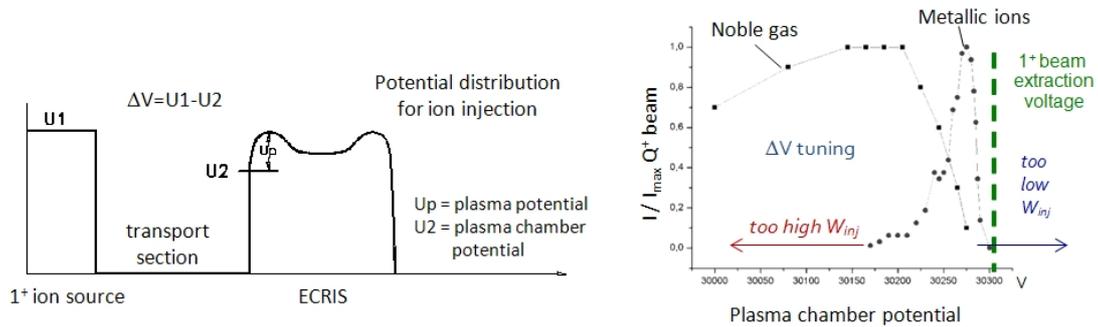

**Fig. 12:** (left) Potential distribution for the primary $1^+$ ion source and the charge-breeding ECRIS. (right) Normalized extracted charge-bred beam as function of the high-voltage potential of the charge breeder for two different primary ion sources. The noble gases were produced in a plasma source with a certain plasma potential, while the metallic ions were produced in a surface ionizer (0 V plasma potential); thus the different optimal potential of the breeder.

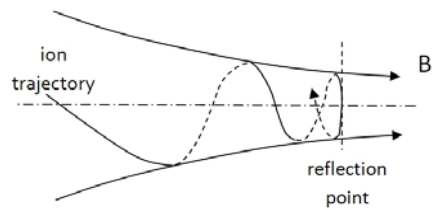

**Fig. 13:** The magnetic-mirror effect showing how a charged particle is reflected in a high B field region. To conserve the magnetic moment μ, the cyclotron frequency is increased while the cyclotron radius is decreased. In turn, the longitudinal momentum is reduced in order to conserve the total kinetic energy.

An ECRIS charge breeder delivers beams with $A/Q$ values between 3 and 9 and an energy spread of a few eV. Breeding efficiencies from a few up to 20% have been reported [26]. Apart from the radioactive beam, many other stable elements originating from the charge breeder are extracted, with a total electric current of around 100 µA. These stable ions originate either from the support gas (often He, O, Ne or Ar) needed to supply the plasma with a permanent source of cold electrons, or from the plasma chamber. Electrons and ions escaping the multi-pole confinement can sputter chamber material into the plasma or desorb ions that have been implanted from previous runs (the so-called memory effect). To limit the $A/Q$ contamination, several techniques related to the plasma chamber cleaning exist [27], and the use of isotopically enriched support gas is recommended.

The normal operation for an ECRIS is CW mode, meaning that the highly charged ions leak out from the plasma towards the extraction electrode. Likewise, the ion injection is also continuous. An ECRIS can, however, be operated in a different mode, the so-called afterglow mode, where the ions are extracted in a pulse coinciding with the switching off of the RF heating. The hot plasma electrons, strongly bound to the magnetic field lines, leave the plasma at the same moment as the RF heating is switched off. The ions, partly confined by the negative space charge of the electrons, are therefore expulsed in a short pulse by Coulomb repulsion. The pulse is a few milliseconds long and shown for $Pb^{27+}$ in Fig. 13 (left). An ECRIS charge breeder operated in afterglow mode can consequently collect ions from a CW ion source, charge breed them and confine them for some 100 ms, and thereafter extract them in a compressed pulse. This has been demonstrated with Rb ions and the general result is illustrated in Fig. 14 (right). The drawback of this mode, also known as ECRIT mode [28], is a reduced transmission efficiency compared with CW operation.

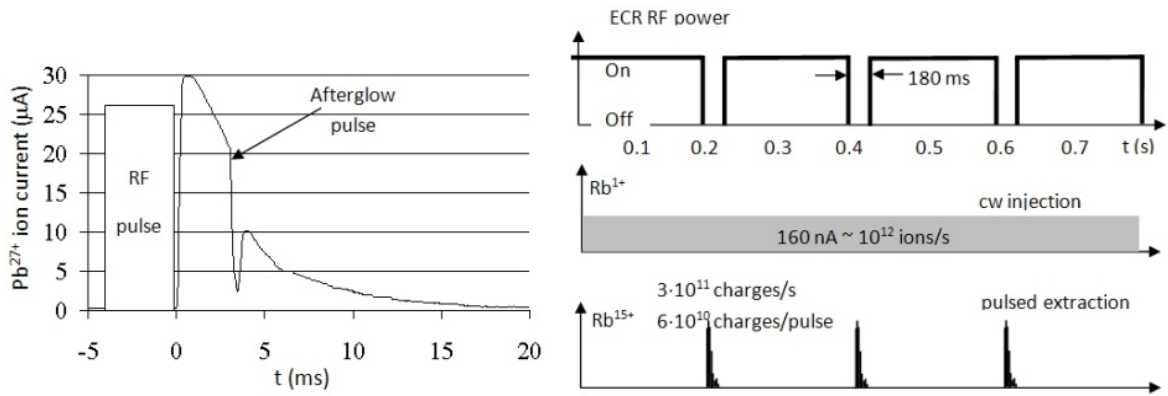

**Fig. 14:** (left) $Pb^{27+}$ afterglow pulse occurring after the RF heating is switched off. Based on [21]. (right) Charge breeding of $Rb^{1+}$ to $15^+$ using the afterglow mode. The pulsed extraction is suited for linacs operated in pulsed mode. Based on [28].

As demonstrated, a charge-breeding ECRIS has to operate with frequency > 10 GHz, and therefore with a magnetic field strength of a couple of teslas. The charge breeder follows the same magnetic field relations as for a high-$Q$, high-current ECRIS, that is, $B_{inj}/B_{ecr} \sim 4$, $B_{ext}/B_{ecr} \sim 2$, $B_{min}/B_{ecr} \sim 0.8$, $B_{rad}/B_{ecr} > 2$ and $B_{ext}/B_{rad} < 0.9$, where $B_{ecr}$ is the magnetic field at ECR resonance, $B_{inj}$ ($B_{ext}$) the field maximum at injection (extraction) side, $B_{rad}$ the radial $B$ field of the sextupole at the plasma chamber wall and $B_{min}$ the minimum $B$ field between the magnetic mirrors. Characteristic for an ECRIS charge breeder is the open end at the injection side, at which the gas and RF injections are usually located for a normal ECRIS. Instead, at this place, a cylindrical tube, connected to a low electrical potential relative to the plasma chamber, guides the ions into the breeder (see Fig. 10). The highest breeding performance is presently achieved by breeders with radial RF injection in the centre of the chamber as a symmetric $B$ field is maintained at the injection region, therefore avoiding beam trajectory deflection of the injected ions, which can be important for light ions [29]. A good pumping conductance is essential to reduce excessive amounts of residual gas ions from the breeder.

The layout of an operational ECRIS charge breeder for radioactive ions is shown in Fig. 15. The use of a combined magnetic and electrostatic analyser will be discussed in section 7.2. The separation of the breeding stage from the production of the $1^+$ ions in the primary ion source means that the breeder is accessible with limited radio-protection restrictions.

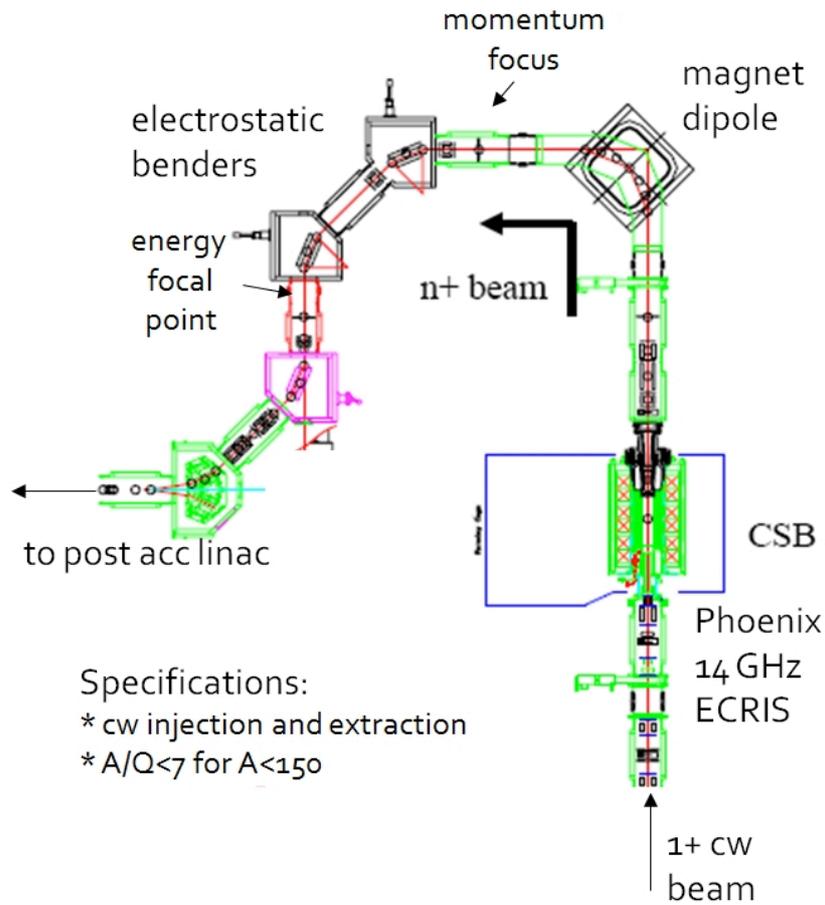

**Fig. 15:** The charge breeder for the ISAC post-accelerator facility at TRIUMF [30]. The $1^+$ beam is charge-bred in a 14.5 GHz ECRIS. The cw extracted beam is separated in a combined magnetic and electrostatic analyser before being injected into a radiofrequency quadrupole (RFQ) cavity and the consecutive accelerating cavities.

## 6  EBIS/T-based charge breeders

By tradition, an EBIS has a long trapping region (>50 cm) with a limited $j_e$ (<1000 A cm$^{-2}$) compressed by an extended solenoid, in most cases superconducting. An EBIT, on the other hand, is based around a pair of Helmholtz coils, has a very small electron-beam diameter $r_{ebeam}$ and therefore has high $j_e$ (>1000 A cm$^{-2}$) owing to the high compression, and a short trapping region (~10 cm), from which usually no ions are extracted. These differences have consequences for the charge-breeding performance, as will be pointed out. For some time now, the distinction between EBIS and EBIT has not been that clear-cut; in fact, the machine types are approaching each other, and EBIT can also provide ions to external set-ups.

The main elements of an EBIS/T, such as electron gun, trapping tubes, electron collector, extraction electrode and magnetic solenoid, are represented in Fig. 16. The ion injection is performed from the collector side, as well as the ion extraction. Once injected, the ions are trapped in a magneto-electrostatic trap: radially by the electron-beam space-charge potential and the magnetic field, and axially by electrostatic potential barriers applied on the trapping drift tubes. The ionization to higher charge states is obtained by bombardment from the fast, dense monoenergetic electron beam generated by the electron gun.

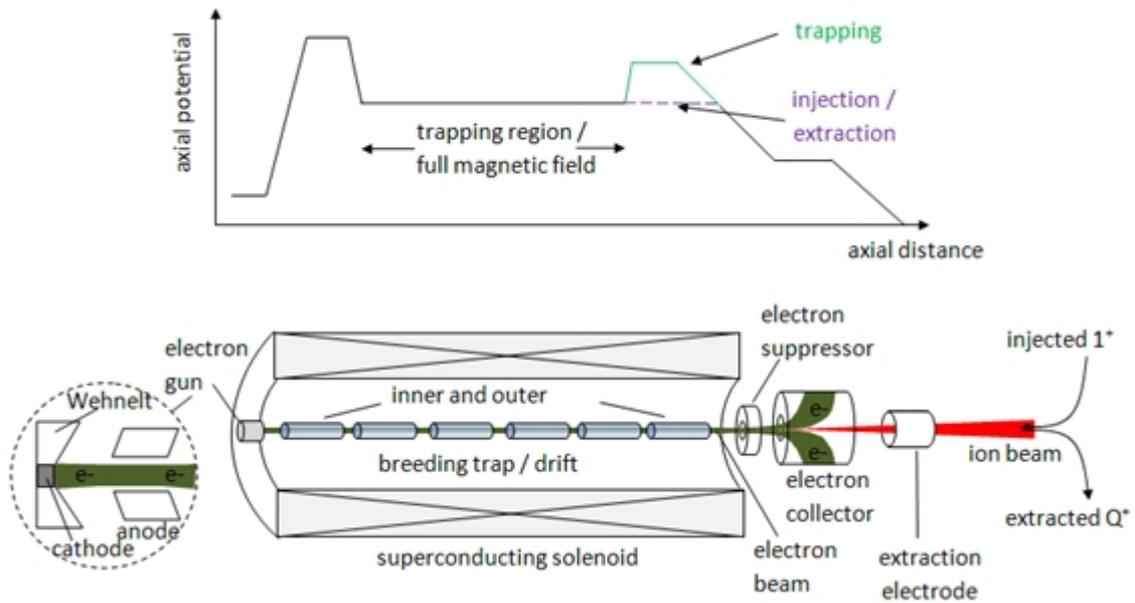

**Fig. 16:** Schematic picture of an EBIS/T. The ion injection and extraction are performed from the collector side using an electrostatic kicker to direct the injected ions from the ISOL system into the breeder and the extracted ions to the succeeding mass separator.

To reach a certain charge state, either the electron current density or the charge-breeding time has be adjusted, as given by Eq. (3). The electron current density, typically varying between 50 and 5000 A cm$^{-2}$, is less convenient to change compared to the breeding time. It is important to have the possibility to select a specific charge state in order to avoid superimposed stable beam contaminants from the breeder, as will be discussed section 7.2. In Fig. 17 two extracted charge spectra of Cs with different breeding times are shown. For a breeding time of 78 ms, the charge-state distribution peaks at 29$^+$; while for 158 ms, the maximum number of ions is found at 32$^+$. In general, one would like to keep the breeding time as short as possible, and therefore one profits from a high total electron current $I_e$ as $j_e = I_e/(\pi r_{ebeam}^2)$.

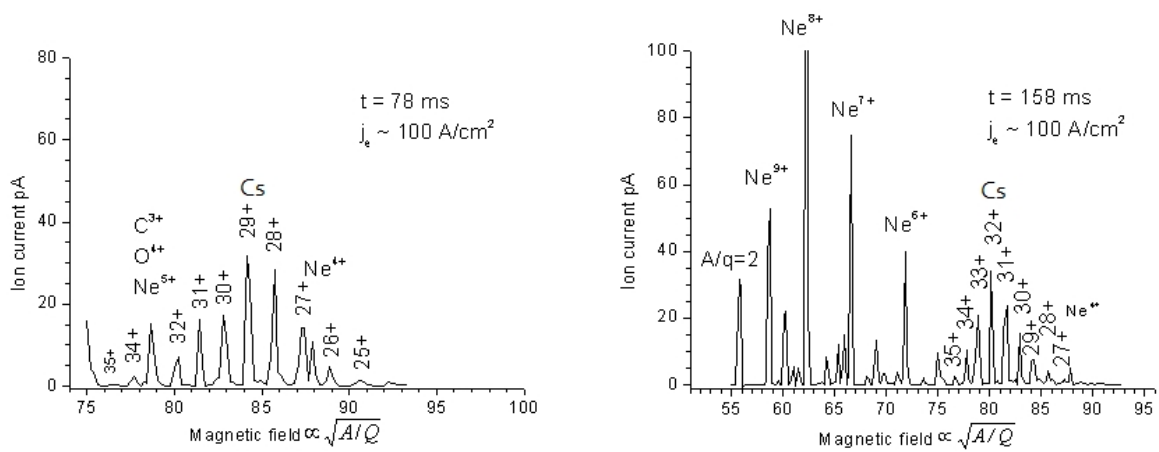

**Fig 17:** Extracted charge-state distributions of Cs for two different breeding times. In both cases, the electron current density was approximately 100 A cm$^{-2}$. The right spectrum is performed with a wider scan so several residual gas peaks are present; most dominant are the Ne peaks originating from buffer gas diffusing from the preceding Penning trap (see section 7.1).

When the ions are trapped inside the breeder, it is desired that they reside within the electron beam in order to undergo stepwise ionization. If they spend part of their time outside the electron beam, the effective electron current density will be lower ($j_e$(effective) = $j_e T_{inside}/T_{total}$) and the charge-breeding time subsequently increases. The ion injection into the EBIS/T is therefore critical, and a maximum overlap between the ion beam and the electron beam is essential. Figure 18 illustrates different ion injection cases, the middle one being ideal, and the right one being completely flawed. The ions will never cross the electron beam in the latter situation and will therefore remain as $1^+$. The left situation, although not ideal but acceptable if a prolonged breeding time is tolerated, will yield charge-bred ions as they traverse the electron beam intermittently and will therefore be further ionized. For each ionization step, the potential depth from electron-beam space charge seen by the ion will increase and, owing to conservation of energy, the ions will be attracted to the beam axis and the overlap therefore increase.

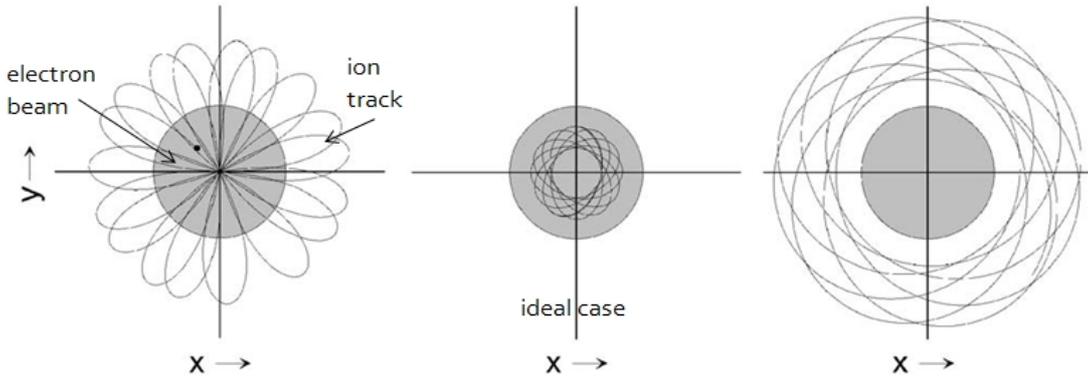

**Fig 18:** Ion tracks for different injection scenarios. To the left the ions occasionally traverse the electron beam and will at some point become ionized to higher charge states, while this never happens in the case to the right. The ideal situation is a complete overlap between the $1^+$ ion trajectory and the electron beam, as shown in the middle. Figure modified from [32].

The transverse phase-space acceptance of the electron beam can be estimated using Eq. (11) [31]. This formula gives the normalized phase-space area within which ions have to fit to be fully trapped within the electron beam when entering a non-space-charge-neutralized EBIS/T:

$$\alpha = \pi\gamma \frac{r_{ebeam}}{c} \sqrt{\frac{Q_{inj}e}{m_{ion}}} \left( Br_{ebeam} \sqrt{\frac{Q_{inj}e}{m_{ion}}} + \sqrt{\frac{Q_{inj}eB^2 r_{ebeam}^2}{4m_{ion}} + \frac{\rho_l}{2\pi\varepsilon_0}} \right) \quad (11)$$

Here the charge per metre of the electron beam is denoted by $\rho_l$ and $\gamma$ represents the relativistic gamma factor. Examining the formula, one finds that two terms originate from the magnetic field, while the second term under the square root is related to the electron-beam space charge. A high electron-beam current results in a large acceptance. Furthermore, an electron beam being space-charge-neutralized by ions has a smaller acceptance compared to an empty beam, and its acceptance is governed by the solenoid magnetic field.[5] As expected, the acceptance is strongly dependent on the electron-beam radius, meaning that traditional EBIT devices with very small beam radii may have difficulties trapping the ions.

There are two schemes to introduce the external ions into the EBIS/T: continuous injection, also called 'accu-mode'; and pulsed injection. In the former, the ions are continuously injected over the outer barrier, and during the round trip inside the trapping region they are ionized to a higher charge state and subsequently trapped (see Fig. 19). Hence, in order to succeed with this injection mode, the ions have first of all to be within the acceptance of the electron beam, as discussed above. In addition, since there are no dissipative forces present (a trapping region with only a few ions confined is assumed), a successive ionization is mandatory, as $1^+$ ions can escape again from the trapping region over the outer barrier. Singly charged ions ionized to $2^+$, on the other hand, will see a doubled outer barrier potential.

---

[5] The partial neutralization of the electron beam by ions is not only harmful. In fact, the compensating ions act as a well-defined Coulomb target for the injected ion and can improve the slowing-down and trapping process, particularly for continuous over-the-barrier injection [33].

Equation (3) gives the average time $\tau_{1\to 2}$ it takes for a singly charged ion to lose one electron, that is $\tau_{1\to 2} = e/(j_e \sigma_{1\to 2})$. The probability $P$ for the ionization process to have taken place within a time $\tau_{ebeam}$ spent inside the electron beam is given by $P = 1 - \exp(-\tau_{ebeam}/\tau_{1\to 2})$. Using the example of $^{55}Fe^+$ with $\sigma_{1\to 2} \sim 10^{-17}$ cm$^2$ being injected into an electron beam with density $j_e = 200$ A cm$^{-2}$, and requiring that $P > 0.5$, a time within the electron beam exceeding 55 μs is required. To increase the probability, the spontaneous solution is to reduce the ion injection energy to increase the round-trip flight time. The magnetic-mirror effect, however, sets a limitation to this, as the ions will be reflected while entering over the outer barrier if the energy is too low, as illustrated in Fig. 12.

To obtain a high trapping efficiency for a continuously injected beam, the trapping region should be long, the electron-beam radius large and the electron current density high. Unfortunately, the two last requirements are in contradiction, as $I_e = j_e \pi r_{ebeam}^2$. In general, a careful tuning of the injection energy is required (within ~10 eV), as well as a small energy spread and transverse emittance of the injected beam in order to obtain an acceptable injection and trapping efficiency. A more efficient injection method is pulsed injection.

The pulsed injection scheme requires a fast pulsing of the outer barrier and a pre-bunching of the beam from the ISOL system. The pulse length has to be shorter than the round-trip time inside the trapping region. The pre-bunching will be dealt with in section 7.1. Figure 20 (top) demonstrates the pulsed injection process with the quick ramping of the outer barrier. To assure an immersed injection into the electron beam, the ion injection energy should ideally be less than the potential depth of the electron beam, given by

$$U_{well} = \frac{I_e}{4\pi\varepsilon_0}\sqrt{\frac{m_e}{2eU_e}} \qquad (12)$$

and illustrated in Fig. 20 (bottom). The notation $U_e$ represents the electron-beam voltage in the trapping region, and a typical value for the electron-beam well is 100 V. For an element with $A = 55$ injected into a 0.5 m long EBIS/T, this corresponds to a pulse length shorter than 50 μs. In reality, somewhat higher injection energy is often used to reduce the magnetic mirror reflection at the entrance at the cost of a longer breeding time (due to the reduced effective electron current density). The pulsed injection method is relatively flexible and has a higher transverse and energy-spread acceptance compared to the continuous injection. The major drawback is the need for pulse formation prior to the breeder.

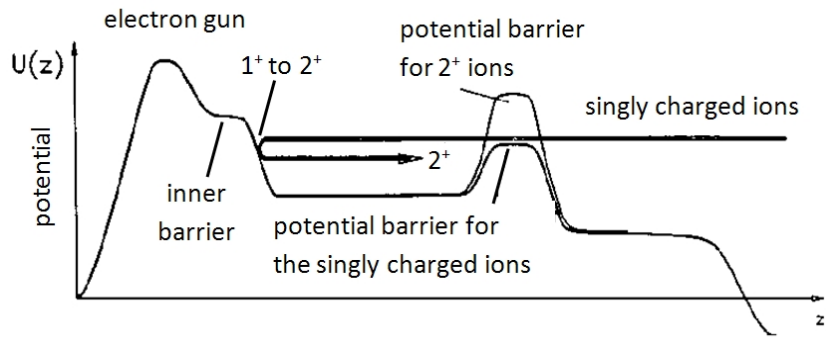

**Fig. 19:** Continuous ion injection into an EBIS/T. The ions are injected with a low velocity over the outer barrier, and inside the trapping region they must undergo ionization to $2^+$ or higher within one round trip to get trapped.

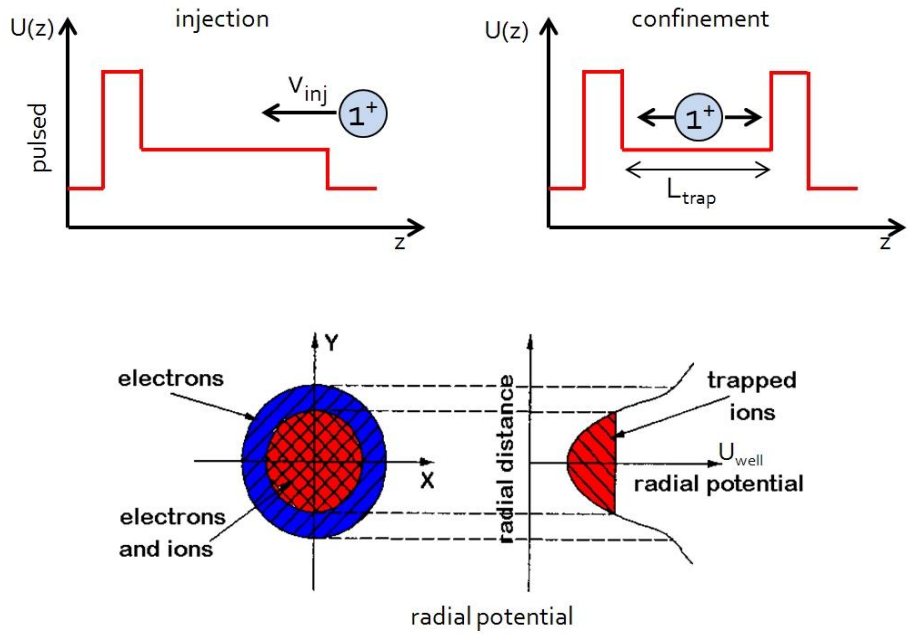

**Fig. 20:** (top) Pulsed injection scheme. The pulse length and the ramping time of the outer barrier are shorter than the round trip within the trapping region. (bottom) The radial electron-beam well and ions injected at the bottom of the well.

The electron-beam energy in a charge breeder is dictated either by the perveance limitation or the desired ion charge states. The perveance, a measure of the space-charge effect in the electron beam, relates the electron current to the electron-beam energy as $P = I_e / U_e^{3/2}$. This is of particular importance at the electron gun, but applies as well for the drift tube and collector regions. As a guideline, the perveance should be below $5 \times 10^{-6}$ perv, which for an electron beam of 1 A corresponds to a minimum electron energy of 3500 V. To reach the desired $A/Q$ value, the electron-beam energy has to be higher than the ionization potential for charge state $Q$ of the element. Particularly difficult are elements along the neutron-rich drip-line due to the excess of neutrons, therefore requiring a high charge state. In Table 3 the ionization potential for different elements along the neutron drip-line are listed assuming that an $A/Q \sim 4$ is requested. Even for the heaviest elements, the ionization potential is modest. Using Lotz's formula, Eq. (4), one can show that the optimal electron-beam energy for ionization is ~2.7 times the ionization potential; thus a maximum electron-beam energy of 10 keV is necessary to cover the full nuclear landscape. Nonetheless, certain dedicated experiments may require much lower $A/Q$ ratios, even fully stripped ions, and then the electron-beam energy should reach 100 keV or higher.

**Table 3**: Ionization potentials for charge state $Q$ for different neutron-rich elements, where the charge state is chosen such that $A/Q \sim 4$.

| Z | A (neutron-rich) | Q | Ionization potential (eV) |
|---|---|---|---|
| 20 | 60 | 15 | 900 |
| 40 | 110 | 27 | 1500 |
| 60 | 161 | 40 | 2800 |
| 80 | 210 | 52 | 3100 |

The extraction of the charge-bred ions from the EBIS/T can be done in three different ways: with passive, fast or slow extraction, shown in Fig. 21. The passive extraction is achieved by abruptly lowering the outer barrier and letting the ions leave the trap with their inherent energy. The total ion energy at extraction is the sum of the injection energy, the energy gained during each ionization step where the potential energy of the ion is increased due to the change in charge, and electron–ion heating and ion–ion cooling processes. Owing to Coulomb scattering among the ions, the momentum space is assumed to be almost isotropic. Heavier ions with higher charge state gain more energy than the lighter ones during the ionization process (even disregarding the electron–ion heating), but the large mass of the ions makes them leave the trap slower. A characteristic passive extraction pulse from a 0.8 m long trapping region ranges from 25 μs (full width at half-maximum, FWHM) for light ions to around 80 μs (FWHM) for very heavy ions ($A/Q = 4$). If an axial electric field is applied during the extraction phase, effectively attained by putting different voltages on the trapping tubes, the ions can be extracted faster (within less than 10 μs) but the energy spread will increase accordingly. Fast extraction is advantageous for injection into synchrotrons, for example.

Finally, by slowly lowering the other barrier, or by slowly increasing the potential of the trapping tubes, the ions can be leaked over the extraction barrier for several 100 μs, or even milliseconds. This is of interest for experiments requiring a low instantaneous particle rate.

The ion throughput rate is determined by the storage capacity of positive charges and the rate at which the injection, breeding and extraction cycle can be repeated. The ion storage capacity is determined by the trap length $L_{\text{trap}}$, the electron-beam current $I_e$, the electron-beam voltage $U_e$ and the electron-beam compensation degree $f$, and the number of charges $N^-$ is given by the expression

$$N^- = \frac{f I_e L_{\text{trap}}}{e} \sqrt{\frac{m_e}{2eU_e}} \; . \tag{13}$$

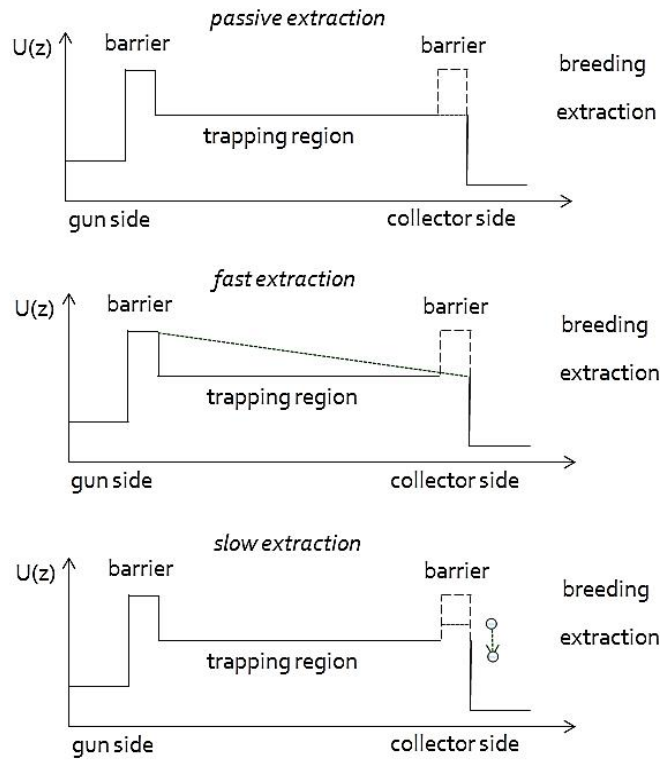

**Fig. 21:** Different ion extraction schemes from an EBIS/T. The fastest extraction is obtained with the second mode, although with a high energy spread. The lowest energy spread, down to a few eV total spread, is reached with slow extraction mode, which also produces long beam pulses.

A long trapping region and high electron current are the main contributors to a high storage capacity. As an example, an electron beam with 0.5 A and 5 keV in a 0.8 m long trap with a 50% compensation degree has a capacity of $3 \times 10^{10}$ charges. With approximately 20% of the beam in the desired charge state, that means that about $10^9$ ions with charge $5^+$ can be extracted from one pulse, and $10^8$ ions with charge $50^+$. This assumes that no evaporative ion cooling [34] using lighter ions is employed, as the light ions will occupy part of the electron space-charge. The ion throughput per second equals the storage capacity divided by the repetition rate, mainly governed by the breeding time. Additional capacity limitation due to a preparatory Penning trap or similar is treated later on in section 7.1.

An example of an RIB facility using an EBIT as breeder is the ReA3 at NSCL/MSU, which is a 3 MeV/$A$ re-accelerator of thermalized projectile fragmentation and fission beams [35]. Its layout is presented in Fig. 22.

The set-up uses a very high-compression electron gun aiming for $j_e > 10^4$ A cm$^{-2}$, meaning that the charge breeding of ions with $Z < 35$ into Ne-like or higher should take less than 10 ms, and the ionization from $1^+$ to $2^+$ occurs within 1 µs. The rapid ionization, in combination with a long trapping region with a variable magnetic field strength in the axial direction, ensures a satisfactory accumulation efficiency for the continuous injection of $1^+$ ions. High ion beam rates (>$10^9$ ions per second) should also be attainable, as the electron current is large and the breeding time very short. The elements in the drift-tube region are at cryogenic temperatures, leading to a good vacuum and low degree of residual gas ions in the extracted beam.

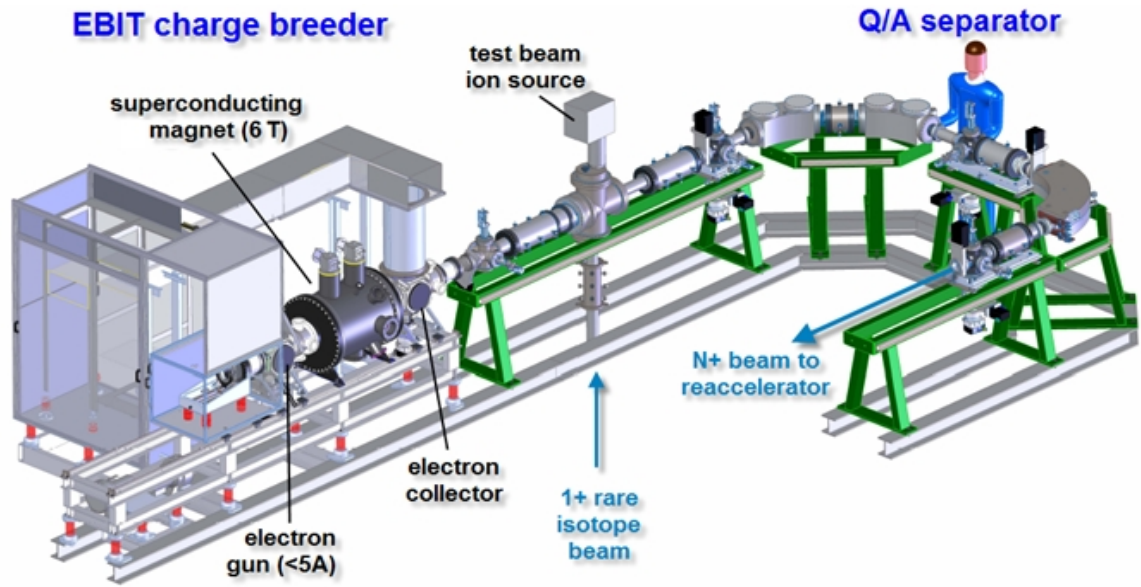

**Fig 22:** The EBIT-based charge breeder for the ReA3 facility. The radioactive $1^+$ ions come from a gas stopper in which they have been cooled to a low emittance and energy spread. The pulsed extracted beam is analysed in an *A/Q* separator before being injected into a superconducting linac.

# 7    Preparatory devices and tricks

## 7.1    Beam preparation in a Penning trap

The performance of the charge breeder can be improved if a beam preparatory device such as buffer-gas-filled RFQ cooler [36] or Penning trap [37] is introduced prior to the breeder. Apart from reducing the transverse beam emittance and the energy spread to a few mm mrad (50 keV, 90%) and to less than 1 eV, respectively, such devices can change the temporal beam structure, that is, bunch the beam.

In addition, the Penning trap approach offers, in the case of low-intensity beams, the extra opportunity of mass purification, as will be explained below.

In a Penning trap, ions are confined in the longitudinal direction by a quadratic electrostatic potential and in the radial direction by a strong magnetic field parallel to the symmetry axis of the electric field (see Fig. 23). The trapped ion performs three independent eigenmotions: one harmonic oscillation in the axial direction ($v_z$) and two oscillations in the radial direction, the reduced cyclotron motion ($v_+$) and the magnetron motion ($v_-$). These frequencies are related as

$$v_\pm = \frac{v_c}{2}\left(1 \pm \sqrt{1 - \frac{2v_z^2}{v_c^2}}\right), \tag{14a}$$

$$v_+ + v_- = v_c = \frac{QeB}{2\pi m_{ion}}. \tag{14b}$$

The cyclotron frequency ($v_c$) depends on the ion mass and the magnetic field strength, while the magnetron motion is almost mass-independent. Typical values for an ion with $A = 30$ in a magnetic field of 3 T, an electrical potential depth of 10 V and a characteristic trap dimension of 15 mm are $v_z = 60$ kHz, $v_- = 1.2$ kHz and $v_c \approx v_+ = 1.5$ MHz.

In a buffer-gas-filled trap, cooling occurs due to the frictional force provided by the gas. The time constant for the cooling is inversely proportional to the frequency of the oscillation. Thus, the axial and reduced cyclotron motions are cooled fast and their amplitudes are consequently reduced. The magnetron motion, on the other hand, is cooled much more slowly and has negative energy radius dependence, that is, the radius increases when the ion is cooled. To prevent ions from hitting the electrode walls, the so-called sideband cooling technique is used [38]. A quadrupolar RF electrical field is applied in the transverse plane at the cyclotron frequency, which couples the magnetron and reduced cyclotron motions. In a trap without a gas, an oscillation between the two motions would occur for a sideband excitation at $v_c$. When gas is added, the cyclotron motion decreases fast, while the magnetron radius increases slowly, but as a result of the coupling both radii are decreased. Since $v_c$ is mass-dependent, the sideband excitation can be used to centre specific ion species as shown below.

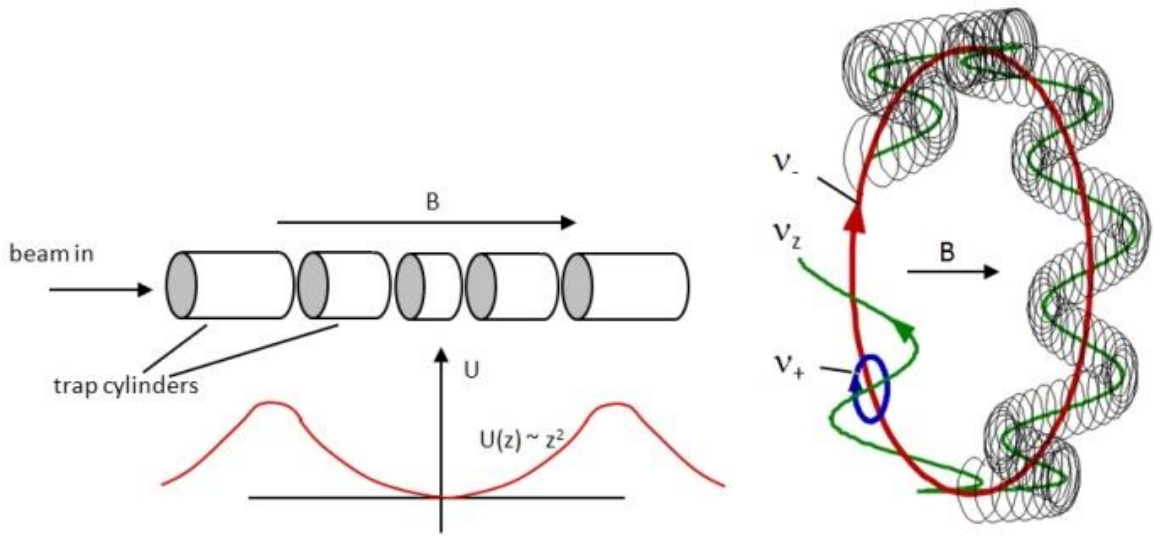

**Fig. 23:** (left) Penning trap electrode stack, with magnetic field and quadratic electrostatic potential. (right) The three ion motions inside a Penning trap: the large-radius magnetron motion with a low frequency $v_-$; the axial motion with a frequency $v_z$; and the fast spinning reduced cyclotron motion with a frequency $v_+$.

To capture the continuous beam from the ISOL system, the ions are slowed down so they have just enough energy to overcome the electrostatic potential wall on the injection side. Because of the presence of the buffer gas, a dissipative force causes the ions to lose energy.

The viscous force can be approximated with $F = -q_{ion}v_{ion}/K$, where $K$ is the ion mobility, which is inversely proportional to the buffer-gas pressure. For example, $^{23}$Na with an energy of 50 eV will lose 25 eV m$^{-1}$ in Ne gas with a pressure of $10^{-3}$ mbar. After a cooling period of some tens of milliseconds, the ions are ejected from the trap as a short bunch (a few microseconds long) when the outer barrier is lowered. The bunch length $\Delta t_{extr}$ and the energy spread $\Delta E_{extr}$ are related by $\Delta E_{extr}\Delta t_{extr} \approx 5$ eV µs.

An example of a preparatory Penning trap is REXTRAP used in front of the REXEBIS charge breeder in the REX-ISOLDE post-accelerator [39]. It accumulates and cools the beam for as long as the charge breeding is on-going inside the EBIS, and, once the EBIS is ready to accept new ions, a bunch is released from the Penning trap. As the extracted beam has a transverse emittance of about 10 mm mrad (30 keV, 90%), and the pulse length is less than 20 µs, the injection into REXEBIS becomes efficient. Normally the Penning trap is operated with a low mass resolution ($m/\Delta m < 300$). The Brillouin limit [40]

$$n_{ion} = \frac{\varepsilon_0 B^2}{2m_{ion}} \quad (15)$$

yields the maximum ion density that can be stored within the trap, and, for a field of a few teslas, around $10^8$ ions with $A = 100$ can be stored per cubic centimetre. In REXTRAP space-charge effects start occurring for more than a several $10^7$ ions per pulse, with an emittance increase and efficiency decrease as a result. This means that, at present, the Penning trap–EBIS concept can provide approximately $10^8 \cdot \eta_{trap\text{-}EBIS}/T_{breeding}$ number of accelerated particles per second, where $\eta_{trap\text{-}EBIS}$ is the combined transmission efficiency for the Penning trap and EBIS, and $T_{breeding}$ the breeding time.

## 7.2  Beam purity

In contrast to other application fields for EBIS/T and ECRIS, a high ion current out of a charge breeder is rarely required. Instead, the intensity of the radioactive beams can be very low, for instance, $10^3$ ions per second corresponding to ~0.2 fA is not exceptional. At such low currents, contaminating beams can easily dominate. In decay studies performed at standard ISOL energies, the stable beam component is of little concern, but for post-accelerated beams it causes problems at the experiment. For instance, the excitation from the contaminations may dominate the direct emission from the experimental target, causing detector overload (pile-up, dead-time and random coincidences) and cross-section normalization problems.

The origin of contamination is either from the ISOL system (radioactive or stable contamination) or from the breeder itself (stable contamination). In the former case, it is almost always a question of isobaric contamination, while from the charge breeder a non-isobaric contamination with a similar $A/Q$ value as the radioactive beam can be present. In the breeder, residual gases are ionized, producing beams of, for instance, C, N and O.

The extracted current is proportional to the partial pressure of the contributing gases, so the good vacuum inside an EBIS/T is advantageous. An $A/Q$ spectrum of an extracted EBIS/T beam is shown in Fig. 24. In an ECRIS, elements sputtered or desorbed from the internal wall surfaces (Al, Fe, etc.) and support-gas ions are abundantly extracted.

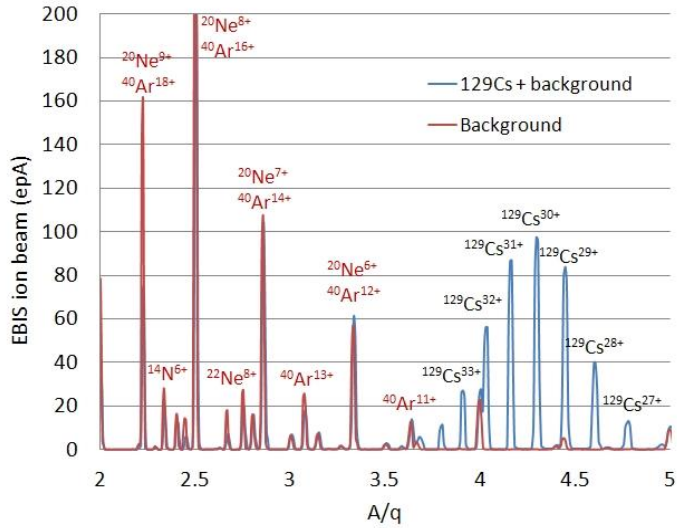

**Fig. 24:** Extracted beams from REXEBIS as a function of A/Q showing residual gas peaks and charge-bred $^{129}$Cs. The blue trace is with and the red trace without $^{129}$Cs being injected. The dominating Ne peaks are due to buffer gas migrating from the Penning trap, while Ar had previously been used as venting gas inside the EBIS.

To suppress the breeder contaminations, a combined electrostatic deflector and magnetic bender is often used in order to attain an achromatic selection system with large energy-spread acceptance [41]. If only a separator magnet, with rigidity $B\rho$, is employed, the velocity spread out of the breeder placed at a potential $U_{extr}$ can cause an ambiguity in the $A/Q$ selection as $B\rho = m_{ion}v_{extr}/q_{ion}$. Nevertheless, with an electrostatic deflector included having an electrical bending field $E_{defl}$ and radius $r_{defl}$, which selects ions according to $E_{defl}r_{def} = 2U_{extr}$, the combined selection becomes $m_{ion}/q_{ion} = (B\rho)^2/(E_{defl}r_{defl})$, and only a single $A/Q$ is transmitted. A typical $A/Q$ resolution of a breeder separator is of the order of a few hundred. Therefore, for example, $^{14}$N$^{6+}$ could possibly be separated from $^{7}$Be$^{3+}$ (required resolution 450), while $^{12}$C$^{6+}$ is difficult to separate from $^{18}$F$^{9+}$ (required resolution 19 200). Owing to the limited resolution, it is imperative to have background-free $A/Q$ regions, as for $^{129}$Cs$^{30+}$ in Fig. 24, and the possibility to tune the ion charge to these regions.

The beam leaving the ISOL system is occasionally superimposed by isobaric contaminants. For instance, the double-magic $^{132}$Sn is overwhelmed by $^{132}$Cs by several orders of magnitude. In the target ion source, several tricks can be applied in order to suppress the contamination, but no universal solution exists. Isobaric selection in the separator is difficult to attain, as a resolution of 5000 to 50 000 is typically needed (see also Table 4).

To combat the isobaric contaminations from the ISOL system, one can use so-called molecular sideband beams. The method makes use of the fact that elements along an isobar have different chemical properties and in most cases do not form similar molecular compounds. Thereby the desired element can be transferred into a clean molecular sideband with a high chemical selectivity. The method is thoroughly treated in [42]. For example, instead of selecting $^{132}$Sn directly from the target and primary ion source, with a substantial $^{132}$Cs contamination superimposed, the molecule $^{132}$Sn$^{34}$S is mass-selected (produced by injection of $^{34}$S$_2$ gas). Cs does not preferably bind as CsS, and is therefore suppressed. The SnS molecule is transferred to the breeder, where it is dissociated and Sn charge-bred in a standard way. The procedure is illustrated in Fig. 25. If some other element of mass 166 leaves the ISOL separator, it can be suppressed in the breeder separator by an appropriate selection of $A/Q$. This two-stage suppression is in general very effective. The transmission efficiency depends on the electronegativity of the desired ion in the molecule, disfavouring radioactive ions with highly electronegative values (particularly group VII elements), since they tend to become neutral (or even negative) when the molecule is broken up and therefore escape the confining potential of the breeder.

Table 4: Typical required mass resolving powers.

|  | Required resolution |
|---|---|
| Neighbouring masses | 250 |
| Molecular ions (e.g. CO from $N_2$) | 500–1000 |
| Isobars (e.g. $^{96}$Sr from $^{96}$Rb) | 5000–50 000 |
| Isomers | $10^5$–$10^6$ |

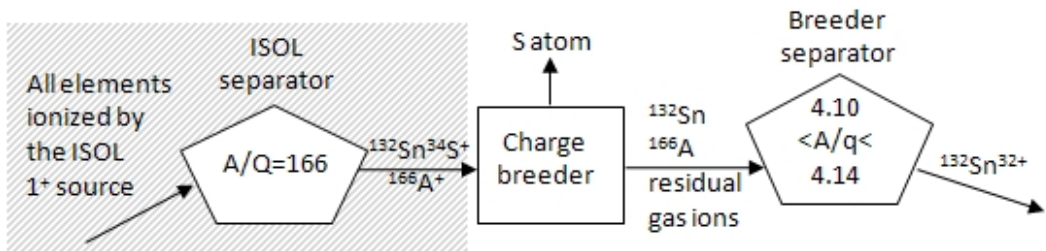

**Fig. 25:** By extracting $^{132}$Sn$^{34}$S from the ISOL system, the contaminating $^{132}$Cs is avoided. The SnS molecule is dissociated in the breeder and Sn is charge-bred and extracted. The $^{166}$A contamination is suppressed in the second separator stage.

Another technique to suppress low-intensity isobaric contaminants from the ISOL system is to make use of the intrinsic mass resolution of a Penning trap. By ion-motion manipulation inside the trap (see Fig. 26), a mass-resolving power of the order of $10^4$ to $10^5$ is achievable in principle. The complete trapped ion cloud is first compressed by a quadrupolar excitation at $\nu_c$ with a large amplitude (the resonance thereby covers all injected elements, including the contaminants) inside the Penning trap for approximately 50 ms. Thereafter the cloud is displaced radially with a dipolar mass-independent excitation at $\nu_-$ for 20 ms. Once offset, a quadrupolar excitation at $\nu_c$ is again applied for 100–200 ms, though this time with a low amplitude and therefore with a narrow sideband resonance, which selectively re-centres the desired ions according to $\nu_c = QeB/(2\pi m_{\mathrm{ion}})$. These can then be extracted from the trap and injected into the breeder in the normal way, while the still displaced contaminants are stopped by a diaphragm. The complete cycle takes up to 400 ms. Owing to space-charge effects, the method is only applicable for less than $10^6$ ions per bunch, including the contaminants [43].

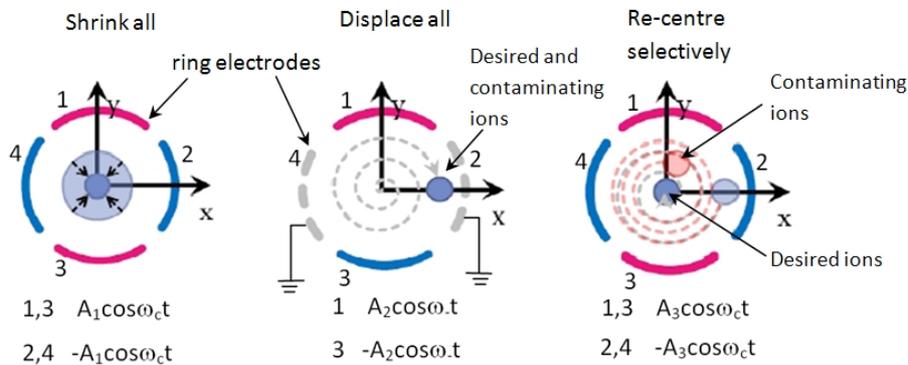

**Fig. 26:** Sequence of RF excitations to achieve isobaric separation in a Penning trap. Axial view of the trap centre. (left) High-amplitude quadrupolar excitation at the true cyclotron frequency to compress the complete ion cloud. (middle) Dipolar excitation at the magnetron frequency to displace the ion cloud outside the acceptance of the extraction diaphragm. (right) Low-amplitude mass-selective quadrupolar excitation at the cyclotron frequency to re-centre the desired ions. Amplitudes $A_1 \gg A_2 > A_3$. Picture adjusted from [44].

# 8 Facilities and future

Charge breeders are not exclusively used as a preparatory stage for injection into a post-accelerator. A few set-ups exist that use EBIT as charge breeder to feed mass-measurement systems, for example TRIUMF's Ion Trap for Atomic and Nuclear science (TITAN) [45]. High-precision mass measurements using traps benefit from the higher charge state as the resolution is given as

$$\frac{m_{ion}}{\Delta m_{ion}} \propto \frac{QeB}{m_{ion}} T_{RF} \sqrt{N} ,\qquad(16)$$

where $T_{RF}$ is the duration of the radiofrequency excitation and $N$ the number of measurement cycles. For short-lived radioactive ions, the advantage becomes evident, as the excitation time may be restricted by the half-life, so instead of several seconds excitation time, a $10^+$ charged ion can be measured with similar accuracy within a fraction of a second. In contrast to the post-accelerator context, these breeders emphasize a high charge state and the amount of ions to breed is limited, of the order of a few hundred, as ideally only a single ion should be injected into the mass-measuring trapping system at each cycle. The efficiency should nonetheless be high, as many of the ions are exotic and produced in small numbers. Additionally, the extracted ions must not be contaminated with stable ions originating from the breeder, as that will impair the mass-measurement process inside the trap.

The use of laser ion sources for the production of highly charged ions has been explored at different laboratories; for example, the production of $Pb^{27+}$ was reported in Ref. [46]. The ionization process is non-resonant and requires high-intensity lasers ($P > 10^{10}$ W cm$^{-2}$) if high charge states are to be attained. Their applicability for charge breeding is, however, expected to be limited, as the efficiency is low for non-solid elements, the transverse beam emittance large (for a Pb beam, 140 mm mrad for 100 kV, 90%) and the energy spread several keV $Q$. The high power requirement also implies that the lasers are pulsed with a limited repetition rate, which negatively affects the efficiency.

The number of charge breeders for RIB facilities have rapidly grown since the year 2000. The present situation, listing operational breeders and breeders in the design or commissioning stage, is illustrated in Fig. 27. All new ISOL facilities involving a post-accelerator, or an advanced Penning trap system for mass measurements, have a breeder of either EBIS/T or ECRIS type.

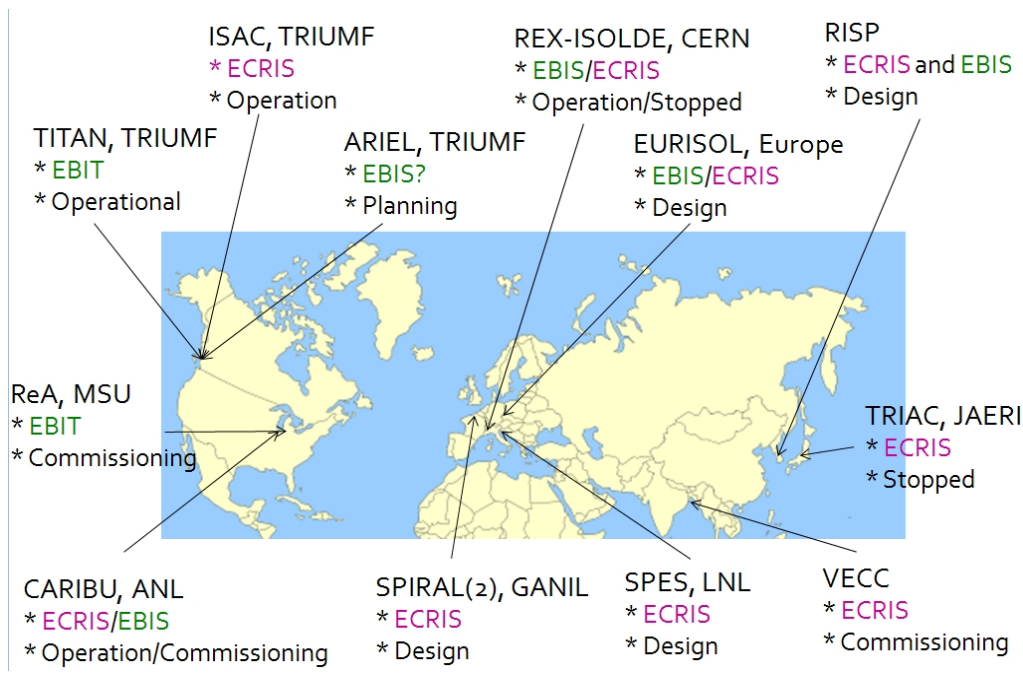

**Fig. 27:** Charge breeders for radioactive ion beams as of spring 2012

From the next-generation RIB facilities, first of all higher beam intensities are to be expected. Accelerated beam intensities exceeding $10^{12}$ particles per second could be expected in an ultimate EURISOL facility [47]. The higher beam intensities implicitly lead to access further out towards the proton and neutron drip-lines, with short nuclear lifetimes as a result. Existing breeding techniques, with parallel operation of EBIS/T for traditional low-intensity experiments and ECRIS breeders for high-current beams, as shown in Fig. 28, will cover these upcoming requirements.

More challenging will be the charge breeding of ions to very high charge states, even fully stripped, for consecutive injection into storage rings [48]. The collection and charge breeding of extreme amounts of $^{6}$He and $^{18}$Ne to be used in beta-beam facilities [49] are also very challenging. The state-of-the-art breeding techniques need to be pushed even further to cover these cases.

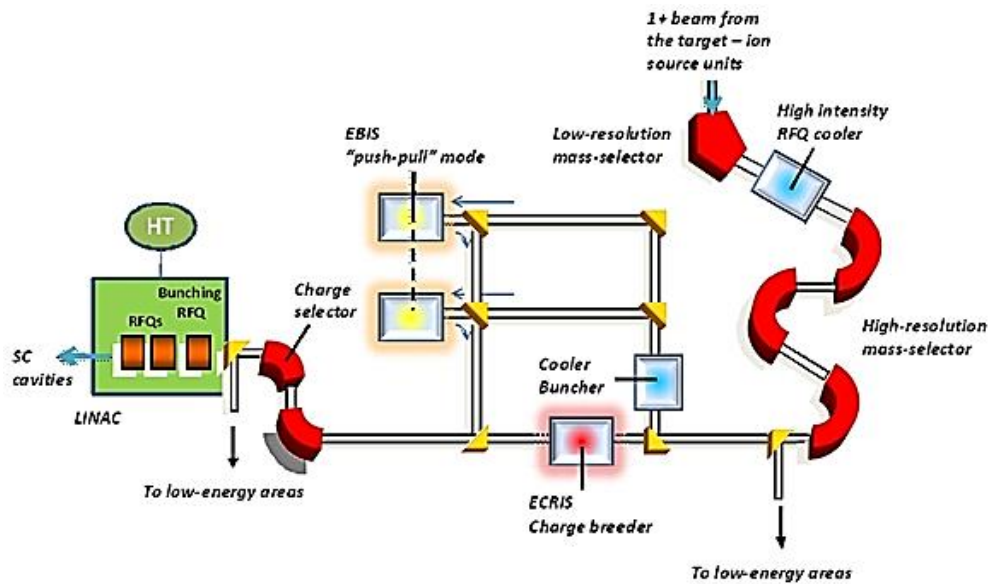

**Fig. 28:** Two main charge-breeding paths in a future high-intensity RIB facility. For exotic low-intensity beams (<$10^8$ ions per second) used for 'standard' nuclear spectroscopy experiments, two EBIS/T operating in push–pull mode are foreseen. For high-intensity RIBs, for example, intended to generate even more neutron-rich nuclei, an ECRIS is used as a charge breeder as the beam purity is of less importance.

## 9    Conclusions

Charge breeding of exotic isotopes produced in ISOL facilities has during the past decade opened up unique possibilities for nuclear and atomic physic experiments at both low and high energies. The higher charge states of the radioactive ions give access to precise mass measurements of very short-lived ions, but, above all, it allows for more compact and cost-effective post-accelerators achieving some MeV/*A* thanks to the lowered mass-to-charge ratio.

The breeding time of some tens to a few hundreds of milliseconds is compatible with the hold-up time in an ISOL target-ion source assembly. As pointed out, the characteristics of the breeding device have a decisive influence on the layout of the post-accelerator, as the time structure of the beam (CW or pulsed extraction) and the *A*/*Q* are determined by the breeder. Furthermore, the level of beam contamination in the extracted charge-bred beam, originating from mainly residual gases in the breeder, is an important aspect.

**Relevant conference Proceedings**

Proceedings of the International Workshop on ECR Ion Sources

Proceedings of the International Symposium on EBIS/T

Proceedings of the International Conference on Radioactive Nuclear Beams (now discontinued)

Proceedings of the International Conference on Electromagnetic Isotope Separators and Techniques Related to Their Applications